\documentclass[a4paper,fleqn]{cas-sc}

\usepackage[numbers]{natbib}

\usepackage{lipsum}
\usepackage{algorithm}
\usepackage{algorithmic}
\usepackage{amsmath}
\usepackage[textwidth=8mm]{todonotes}
\usepackage[normalem]{ulem}

\usepackage{lineno}

\def\tsc#1{\csdef{#1}{\textsc{\lowercase{#1}}\xspace}}
\tsc{WGM}
\tsc{QE}

\begin{document}
\let\WriteBookmarks\relax
\def\floatpagepagefraction{1}
\def\textpagefraction{.001}

\shorttitle{ABM of Elephant Crop Raid Dynamics}    

\shortauthors{Anjali Purathekandy, Meera Anna Oommen, Martin Wikelski, Deepak N Subramani}  

\title[mode = title]{An Agent-Based Model of Elephant Crop Raid Dynamics in the Periyar-Agasthyamalai Complex, India}  



%

\author[1]{Anjali Purathekandy}[type=editor, style=chinese, orcid=0009-0002-4257-5489]



\ead{anjalip1@iisc.ac.in}


\credit{Conceptualization, Methodology, Software, Formal analysis, Investigation, Visualization, Writing - Original Draft}
\author[2]{Meera Anna Oommen}[style=chinese]


\ead{meera.anna@gmail.com}

\ead[url]{https://www.dakshin.org/dt_team/meera-anna-oommen/}

\credit{Investigation, Data Curation, Validation}
\author[3,4]{Martin Wikelski}[style=chinese, orcid=0000-0002-9790-7025]


\ead{wikelski@ab.mpg.de}

\ead[url]{https://www.mpg.de/307762/animal-behavior-wikelski/}

\credit{Investigation, Data Curation}
\author[1]{Deepak N Subramani}[style=chinese, orcid=0000-0002-5972-8878]

\cormark[1]


\ead{deepakns@iisc.ac.in}

\ead[url]{https://cds.iisc.ac.in/faculty/deepakns/}

\credit{Conceptualization, Methodology, Formal analysis, Validation, Supervision, Writing - Review and Editing}
\affiliation[1]{organization={Indian Institute of Science},
            addressline={Department of Computational and Data Sciences}, 
            city={Bangalore},
            citysep={}, 
            postcode={560012}, 
            state={Karnataka},
            country={India}}
            
\affiliation[2]{organization={Dakshin Foundation},
            addressline={No.2203, 8th Main, D Block, MCECHS Layout, Sahakar Nagar}, 
            city={Bangalore},
            citysep={}, 
            postcode={560092}, 
            state={Karnataka},
            country={India}}

\affiliation[3]{organization={Department of Migration},
            addressline={Max Planck Institute of Animal Behavior}, 
            city={Konstanz},
            country={Germany}}

\affiliation[4]{organization={Centre for the Advanced Study of Collective Behaviour},
            addressline={University of Konstanz}, 
            city={Konstanz},
            country={Germany}}
\cortext[1]{Corresponding author}


\begin{abstract}
Human-wildlife conflict challenges conservation worldwide, which requires innovative management solutions. We developed a prototype Agent-Based Model (ABM) to simulate interactions between humans and solitary bull Asian elephants in the Periyar-Agasthyamalai complex of the Western Ghats in Kerala, India. The main challenges were the complex behavior of elephants and insufficient movement data from the region. Using literature, expert insights, and field surveys, we created a prototype behavior model that incorporates crop habituation, thermoregulation, and aggression. We designed a four-step calibration method to adapt relocation data from radio-tagged elephants in Indonesia to model elephant movements in the model domain. The ABM's structure, including the assumptions, submodels, and data usage are detailed following the Overview, Design concepts, Details protocol. The ABM simulates various food availability scenarios to study elephant behavior and environmental impact on space use and conflict patterns. The results indicate that the wet months increase conflict and thermoregulation significantly influences elephant movements and crop raiding. Starvation and crop habituation intensify these patterns. This prototype ABM is an initial model that offers information on the development of a decision support system in wildlife management and will be further enhanced with layers of complexity and subtlety across various dimensions. Access the ABM at \url{https://github.com/quest-lab-iisc/abm-elephant-project}.
\end{abstract}



\begin{highlights}
\item Developed a new prototype ABM that incorporates movement models, behavioral models and crop raid dynamics informed from field data (either direct or substitute) and literature.
\item Movement model calibrated through a four-step process using radio-tagged elephant relocation data from Indonesia.
\item Behavioral insights, conflict validation, and crop patterns in the model domain informed from a questionnaire based field survey in the Periyar-Agasthyamalai Complex.
\item Crop habituation and foraging efficiency emerges as a driver for crop raids even in seasons with food abundance in the forest
\item Thermoregulation emerges as a counter-intuitive factor for reducing conflict in dry/hot months
\end{highlights}


\maketitle

\section{Introduction}\label{Introduction}

Communities near forest fringes face significant human-wildlife conflicts, worsened by habitat degradation and competition for resources such as space, food, and water \citep{Goumas2020}. Marginalized communities are the worst affected, suffering human casualties, crop and livestock losses, and property damage \citep{Gulati2021}. Beyond direct costs, conflicts incur hidden costs including food and financial insecurity, disrupted lives, and expenses from mitigation strategies \citep{Bond2018,Gulati2021}. Wildlife also face long-term survival threats, increased stress levels, and food scarcity as a result of these conflicts \citep{Karanth2010,Pokharel2019}.

The primary models for studying human-wildlife conflict are Equation-Based Models (EBM), Game-Theoretic Models (GTM), and Agent-Based Models (ABM) \citep{Neil2020}. EBMs simplify complex systems for broad understanding (e.g., predator-prey models \citep{Beddington1975,Gard1976}). GTMs, like green security games, help mitigate poaching and strategize ranger actions \citep{Fang2017,Xu2017}. They improve decision-making when integrated with EBMs or ABMs. On the other hand, ABMs provide a flexible framework that captures the heterogeneity of complex systems more effectively than EBMs, making them ideal for modeling individual differences, social structures, and decision-making processes \citep{Neil2020,Mamboleo2021}. They excel in addressing the diverse spatio-temporal drivers of human-wildlife conflict, including environmental factors such as habitat loss, forest fragmentation, and resource scarcity, and behavioral factors such as wildlife's food habituation and human cultural perceptions \citep{Anwar2015,Vinod2019,Donaldson2012,Marley2017,Marley2019}. The temporal dynamics of these interactions is influenced by human disturbance and seasonality \citep{Kalyanasundaram2013,Chandima2022}. ABMs effectively characterize these complex interactions, serving as a virtual laboratory to explore and analyze critical dynamics in space and time \citep{Mamboleo2021}.

Human-elephant conflict occurs when human and elephant needs clash, leading to antagonistic interactions and co-existence challenges. Such conflicts often arise from competition over resources such as food, water, and space, triggered by natural or anthropogenic factors. These conflicts, while not implying hostility, indicate significant co-existence difficulties. Our prototype ABM specifically addresses crop and infrastructure damage as conflict scenarios.

ABMs have been extensively deployed to investigate the complexities of human-wildlife conflict \citep{Musiani2010,Fernando2013,Burton2013,Marley2017,Marley2019}. In addressing the human-elephant conflict, a predictive ABM was formulated to examine the fluctuations in demographic rates of elephant populations, attributable to spatio-temporal vegetational changes \citep{Boult2018}. The median value of the Normalized Difference Vegetation Index (NDVI) served as a proxy for food availability within the model. Despite the model's ability to reflect demographic data variability, its reliance on a uniform food value across the study landscape does not adequately represent the intricate behavioral dynamics of both the environment and the elephants. Furthermore, a prototype ABM for poaching mitigation was developed within a simulated protected area, testing various poaching and law enforcement strategies \citep{Neil2020}, thereby underscoring the potential of ABMs as instrumental management tools in multifaceted scenarios. ABMs have also been utilized to elucidate both anthropogenic and natural factors influencing human-elephant interactions amid resource constraints in the Bunda district of Serengeti National Park, Tanzania \citep{Mamboleo2021}. Additionally, an ABM incorporating a dynamic vegetation model was employed to evaluate the repercussions of climate change on elephant distribution in Kruger National Park, South Africa \citep{Clemen2021}. The spatial dynamics of African Savannah elephants, predicated on the accessibility of essential resources such as food, water, and shade, were modeled through a spatially-explicit ABM \citep{Diaz2021}. This model investigated various movement dynamics including daily net and diel displacements, home range extents, and proximity to permanent water bodies, concluding that water availability critically shapes the spatial behavior of these elephants. Employing data from Gorongosa National Park, Mozambique, a combinatorial optimization technique was integrated into an ABM to delineate elephant movement patterns and scrutinize spatial utilization \citep{Haosen2022}.

Previous studies have not explored the application of ABMs in modeling the behavior of Asian elephants (\textit{Elephas maximus}) and their interactions with humans in areas marked by human habitation and agricultural activities. This paper aims to create a prototype ABM to examine and understand the dynamics of human-elephant conflicts, focusing on crop raiding in Seethathode, Kerala, India, as a specific example. Located on the forest fringe, this area has experienced numerous conflicts between humans and elephants. A comprehensive interdisciplinary research \cite{Oommen_2018} has documented these conflict scenarios and investigated their underlying causes, providing essential data for the development of our ABM.

In an earlier ABM for the African elephant \citep{Mamboleo2021}, the elephant agents would only invade agricultural lands under conditions of food or water scarcity, as indicated in the model. However, this is not the sole motivator for such behavior, as other elements might also drive elephants towards human settlements. For example, Crop habituation might explain the frequent crop raids observed on the forest fringe such as Seethathode \citep{Kalyanasundaram2013,Rohini2015,Dangol2020,Chandima2022}. Even when food is available in the forest, elephants may still raid crops due to a preference for the higher caloric value of farm crops on the outskirts \cite{Sukumar1990}. Importantly, the behavioral traits of elephants, including their aggression and propensity to take risks, significantly influence these raids. Furthermore, the timing of fruit seasons and human activity patterns significantly affect agricultural operations in Seethathode and similar regions \citep{Barnes1991,Kumar2010,Srinivasaiah2012,Jathanna2015,Gaynor2018,Wilson2021}. Elephants, for example, tend to raid farms at night when there is less human activity. In this study, we propose a more detailed conflict model that incorporates these varied factors.

For the development of our model, we prioritized the use of data related to the ecology and behavior of the Asian elephant in the Western Ghats whenever feasible. In cases where such data were scarce, we turned to insights gathered from the same species in different regions, and also drew upon some knowledge from the African elephant when we found such comparisons to be pertinent. Acknowledging the challenge of simplifying the behavior of complex species into mathematical expressions, we depended on expert insights (Sanjeeta Sharma Pokharel, personal communication, April 19, 2024) and conducted parameter sweeps in our model to manage uncertainties. Our ABM is intended to serve as an initial model that incorporates diverse aspects of the conflict with flexibility to enhance complexity in a modular fashion. 

In what follows, in Sect.~\ref{sec:Materials_and_Methods}, we describe the field data and the development of the ABM following the standard Overview, Design concepts, Details (ODD) protocol \citep{GRIMM2006,GRIMM2010,MULLER2013,GRIMM2020}. Next, the experimental setup (Sect.~\ref{experimental_setup}), model validation (Sect.~\ref{Model_validation}), insights into the results (Sect.~\ref{Results}) and discussion of the findings with respect to the literature, including a list of the limitations of the current prototype model (Sect.~\ref{Discussion}) are provided. Finally, we close with concluding remarks and future directions (Sect.~\ref{sec:Conclusions}). 

\section{Materials and methods}\label{sec:Materials_and_Methods}

\subsection{Study area}
\begin{figure*}[!h]
    \centering
    \includegraphics[width=1\linewidth]{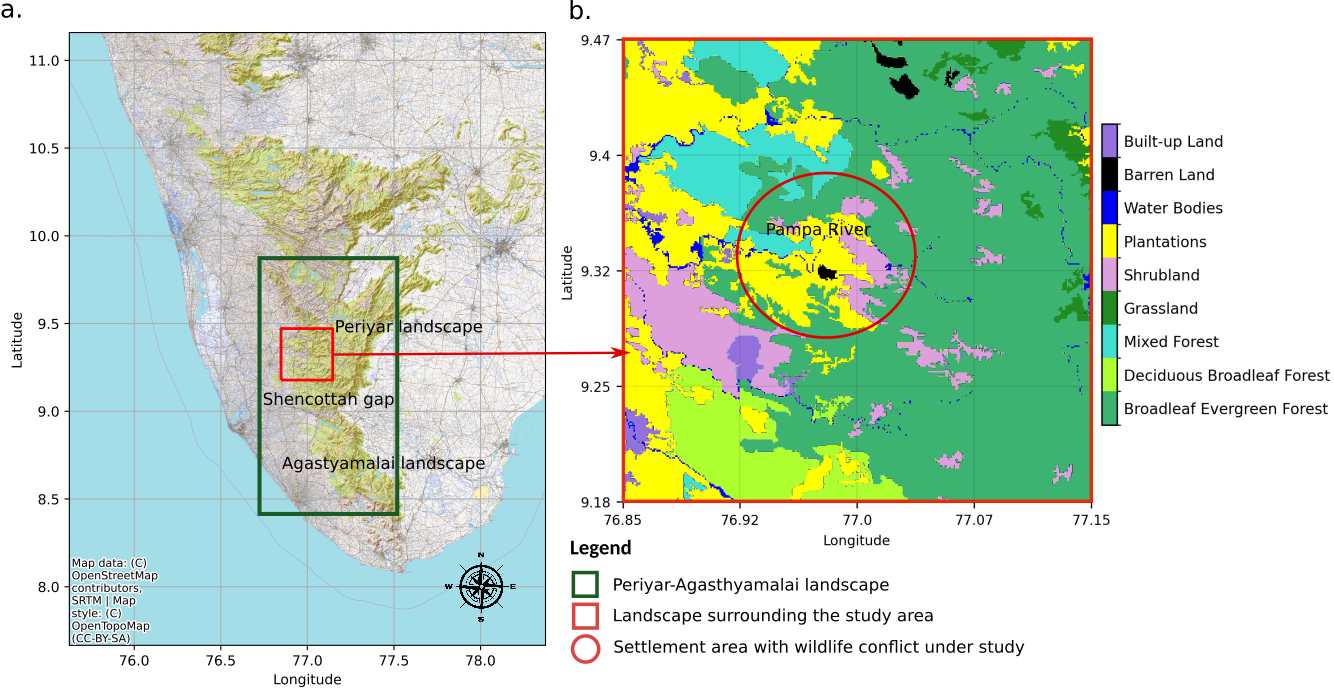}
    \caption{(a) Location of the study area within the Periyar-Agasthyamalai landscape in the southern Western Ghats range. (b) Map of detailed Land Use and Land Cover (LULC) of the study area's landscape.}
    \label{fig:map_detailed}
\end{figure*}

We focus on Seethathode village in Kerala, India, to develop our ABM. The study area (red box in Fig. \ref{fig:map_detailed} (a)) is part of the Periyar – Agasthyamalai landscape of the Western Ghats (green box in Fig. \ref{fig:map_detailed} (a)), which consists of two mountain ranges: the Periyar Hills and the Agasthyamalai Hills. The Western Ghats are a range of mountains that run for approximately 1,600~$km$ parallel to India's western coastline, stretching from Gujarat in the north to Kerala in the south. The topography of the Periyar - Agasthyamalai landscape has a wide range of elevations, ranging from approximately 100 to 1800~$m$ above sea level. The Periyar Hills and the Agasthyamalai Hills are separated by the Shencottah Gap, through which the National Highway 744 and a railway line run. This area has numerous human settlements and commercial plantations that lead to fragmentation of elephant habitats between the Periyar and Agasthyamali forest complexes. This region has been the focus of a multidisciplinary long-term investigation study on human-wildlife conflict on the forest-agriculture fringe \cite{Oommen_2017_a}. This previous study identified ecological and social drivers of conflict and contributed a range of insights and data to the development of our ABM.


\begin{figure*}[!h]
    \centering
    \includegraphics[width=1\linewidth]{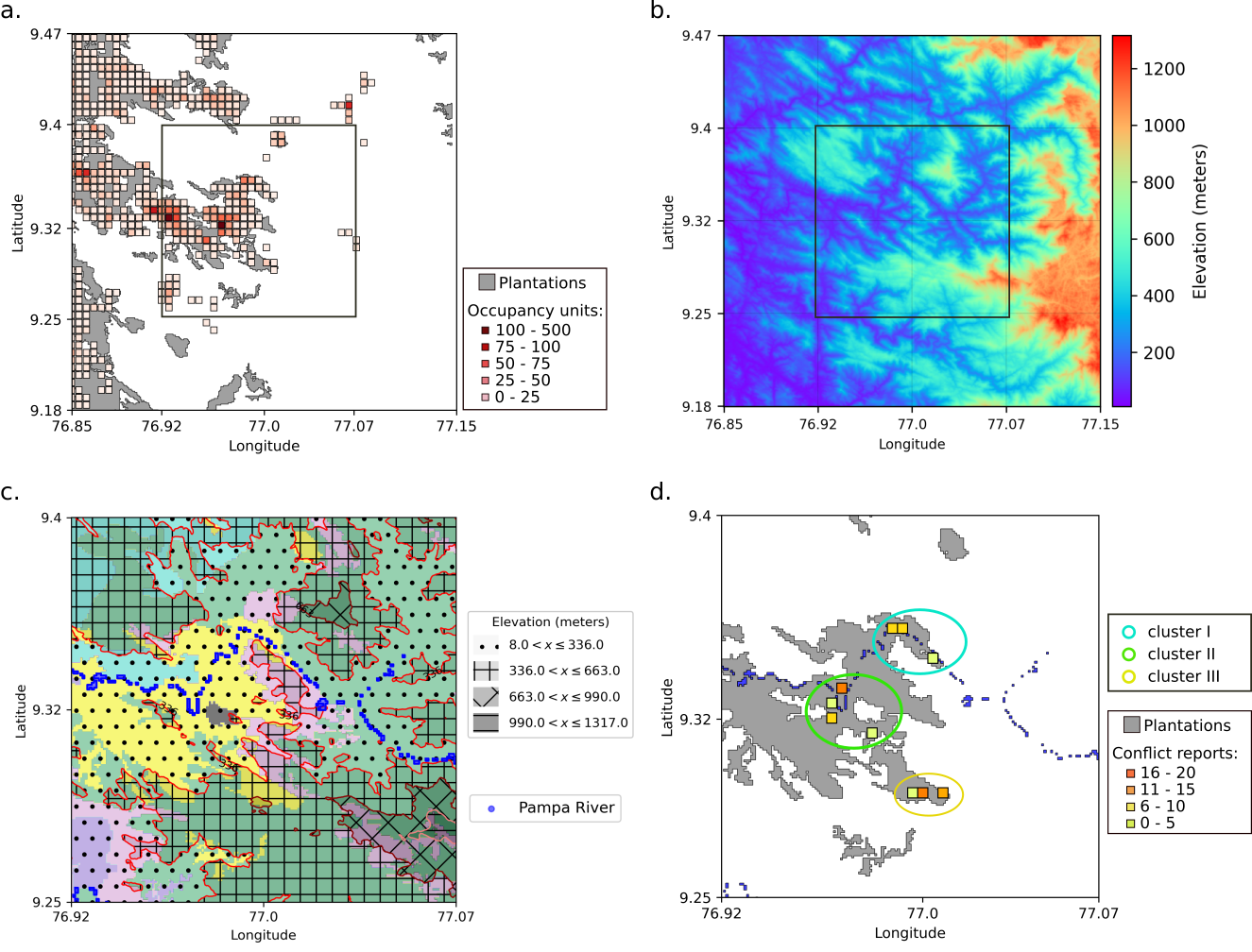}
    \caption{(a) Occupancy/building unit density within the study area. Each cell represents an area of 0.5~$km^2$. where buildings are present. Buildings are concentrated in areas classified as plantations. (b) The Digital Elevation Model (DEM) of the study area. (c) The elevation contour plot against the LULC map. The Pampa River flowing through is highlighted. The most accessible pathways for elephants are the low-lying regions that connect to plantations and human settlements (elevation: $8.0 < x \leq 336.0$). (d) Identified clusters of human-elephant conflict based on reported incidents. Panels (c) and (d) refer to the square-marked regions in the remaining panels, which specifically highlight the disputed settlement area.}
    \label{fig:study_area_detailed}
\end{figure*}

The study area is a square region located between latitudes 9.177$^o$N and 9.471$^o$N and longitudes 76.848$^o$E and 77.146$^o$E (Fig. \ref{fig:map_detailed} (b)). This area was selected to include the forest fringe, the human settlement zone with reported incidents of elephant attacks, and the considerable forest landscape surrounding it. This area is located within the Ranni Forest Division in Pathanamthitta district of Kerala (India), which is part of the biodiversity corridor that connects two large protected area complexes to the north and south. 

Broadleaf forests and plantations are the main land use categories in the study region. Other categories such as grasslands, water bodies, barren land, shrublands, mixed forests, and built-up land occupy only a small portion of the study landscape. Crop cultivation and human settlements are concentrated in areas designated as plantations (Fig. \ref{fig:study_area_detailed} (a)). Human activities are concentrated in the valleys, resulting in a higher population density in these lowland areas. As elevation increases along steep mountain slopes, human activity gradually decreases. The terrain presents inherent challenges to movement due to its steep slopes and rugged topography (Fig. \ref{fig:study_area_detailed} (b)). Humans and elephants traverse the hilly terrain, mainly using relatively flatter and more accessible routes in the valleys or passes for their movement (Fig. \ref{fig:study_area_detailed} (c)). 

The agriculture in the region is mainly rubber cultivation, with various fruit crops such as plantain, jackfruit, and mango scattered throughout. The climate seasons are the southwest monsoon (June to September), the northeast monsoon (October to January) and a dry season (February to May). The region experiences heavy rain during monsoons. The Pampa River flows through the region and is a source of fresh water, especially during the dry season.

\subsection{Field survey for data collection in Seethathode Village}
\label{sec:field_survey_and_data_collection}


Field information on conflict related to the simulation domain was provided by Oommen \cite{Oommen_2017_a}. The proximity of the plantation area to the forest has led to consistent crop raids by various species of wildlife, such as wild boar, monkeys, and deer, which present recurring concerns.

The primary mode of data acquisition involved conducting questionnaire surveys and interviews through in-person visits to residential households as well as a historical exploration of conflict related to elephants in the region by the co-author Oommen \cite{Oommen_2017_b, Oommen_2018}. During these interviews, the research team collected information on human-elephant conflict, human-wildlife conflicts with other animals, land use types, and agricultural practice including crop choice among other information (More details in Appendix \ref{sec:dataAvailability}). The GPS coordinates were also meticulously recorded for each surveyed location, providing precise geospatial references. The total number of sampling locations was 481, of which 386 faced some competition with wildlife. The data provided relate to conflict incidents that occurred before and during 2010. However, no specific timestamp was recorded for when these conflicts occurred. Additionally, the data collected were biased due to limitations such as the limited number of interviews conducted and the reluctance of individuals to disclose sensitive information. However, it was widely acknowledged that conflict, as a recurring issue, can potentially disrupt daily life.

Data obtained from the questionnaire survey indicate mixed opinions about fencing as a strategy to mitigate human-wildlife conflict. Among the respondents, 32\% reported using different types of fencing (such as barbed, electric, or trench), but only 22\% considered it a viable solution overall. It is important to mention that only 12\% of the participants who used fencing expressed their support for its wider implementation, indicating possible limitations on its efficacy. This highlights the need to explore alternative approaches beyond conventional fencing interventions. The survey also revealed a range of suggested solutions, including non-violent deterrents like using tin sheets to scare animals. In addition, respondents called for government intervention, community participation, and extreme measures such as culling of wildlife. Crop raiding was identified as the primary concern among people who reported conflict, followed by concerns about monkey incursions and attacks on domestic animals.

Although conflicts reported by other animal species exhibit a more uniform distribution along the forest fringe, collected data reveal three distinct clusters of human-elephant conflict concentrated within specific regions of the study area (Fig. \ref{fig:study_area_detailed} (d)). The identification of clusters within the conflict data reported was performed using the Density-Based Spatial Clustering of Applications with Noise (DBSCAN) algorithm. DBSCAN classifies data points into clusters based on their proximity and density, effectively distinguishing dense regions as clusters while labeling isolated points as noise. The ability of this algorithm to handle varying cluster shapes and sizes makes it well suited for the analysis of conflict data characterized by complex spatial patterns. These identified clusters are used as a baseline for the validation and calibration of our agent-based model. Other conflict statistics extracted from the data were also used for model calibration [see Sect.~\ref{sec:Calibration_of_the_model}]. 

\subsection{Radio-tagged movement data for Asian elephants}
\label{sec:movement_data}
Radio-tagged movement data of elephants from the Periyar-Agasthyamali complex would have been the ideal and most straightforward data set to build the movement model of Asian elephants in this terrain. Unfortunately, such a detailed movement data, or at least monthly area usage data and daily travel distances in the simulation domain were unavailable. Thus, we used the relocation data from Jambi, Indonesia, to build and calibrate a movement model for elephants in our domain.

Relocation data on Asian elephant movement were obtained from the study: \textit{Elephants Java FZG MPIAB DAMN} (ID - 56232621; accessed on 3 May 2022) from \textit{movebank.org}. This data set consisted of geolocations with timestamps, documenting the movement of a radio-tagged adult female Asian elephant in Jambi, Indonesia, during the period March 2015 to December 2015 at a 5-minute interval (See the Appendix \ref{sec:dataAvailability} on Data Availability). The process of obtaining permits, the transfer of tags from South Africa to Indonesia, the high-risk field work, and the management of the data were carried out by the co-author Wikelski. This information was used to calibrate the model parameters so that the elephant agent's movement statistics matched the relocation data (Sect.~\ref{sec:Calibration_of_the_model}). 

\section{Model description}
\label{Model_description}

In what follows, we describe the details of the model adhering to the established Overview, Design concepts, and Details (ODD) protocol for ABMs \citep{GRIMM2006,GRIMM2010,MULLER2013,GRIMM2020}. The model is developed using Mesa (version 1.1.1) \citep{python-mesa-2020}, a Python-based ABM framework (version 3.9.13).

\subsection{Purpose and patterns}
\paragraph{Purpose}
The ABM is developed as a computational tool to study the spatio-temporal aspects of the human-elephant conflict problem, defining crop raid incidents as conflicts, in the forest fringe areas characterized by human settlement and crop cultivation. 
Specifically, the model is intended to provide insight into how the different drivers of crop raid incidents (both behavioral and environmental) on a microscale replicate different patterns of crop raid incidents on a macroscale. As a by-product the space use characteristics on Asian elephants in the western ghats is also studied.
The overarching objective of this development is to establish a foundational model for an advanced decision support system, designed to predict conflict scenarios in response to different mitigation strategies and evolving environmental conditions.
\paragraph{Patterns}
The usefulness of the model was evaluated on the basis of the reproducibility of the following patterns.

\begin{enumerate}

    \item Emergent space-use pattern of the elephant that is governed by the elephant's cognition model, resource availability in the environment, and the behavioral response of the elephant to the available resources.
    
    \item Emergent crop raid and infrastructure damage incidents that are governed by the distribution of agents, characteristics of the environment, the availability of resources, and behavior model of the elephant agents.

\end{enumerate}
\subsection{Entities, state variables and scales}
\paragraph{Entities}


There are two entities present in the model: the elephant agent and the environment (discretized into landscape cells). In the present ABM, only crop raid incidents are considered and a direct human agent-elephant agent interaction is not included. As such, human agents are absent in this prototype ABM.

\textbf{Elephant agents} Two distinct social structures are found within the populations of Asian elephants, namely matriarchal herds and solitary bulls. The matriarchal herd consists of adult females, and subadult/juvenile males and females. Subadult males disperse from the group after puberty and are considered as solitary bulls. There are also associations between these groups, which depend on factors such as the season, resource availability, competition for food and mates \citep{Silva2012,Nandini2017,Nandini2018}. Pubertal and adult male elephants were found to take more risks compared to female-led matriarchal herds \citep{sukumar1988a}. Although the pattern of crop raiding varied from place to place, male elephants exhibited more aggression in terms of the frequency of crop raid incidents and per capita damage \citep{sukumar1988a,Baskaran1995,Williams2001,Sampath2011}. The present prototype ABM focuses solely on individual bulls due to their higher propensity to take risks and their greater potential for damage. Our goal is to get a first model that provides insights in a limited setting now, but with full capability of future expansion by adding layers of complexity. To extend the prototype ABM to incorporate matriarchal herds in the future, the cognition parameters must be updated and a new model for the association between individuals in the group must be introduced.

\begin{figure*}[h!]
\centering
\includegraphics[width=\linewidth]{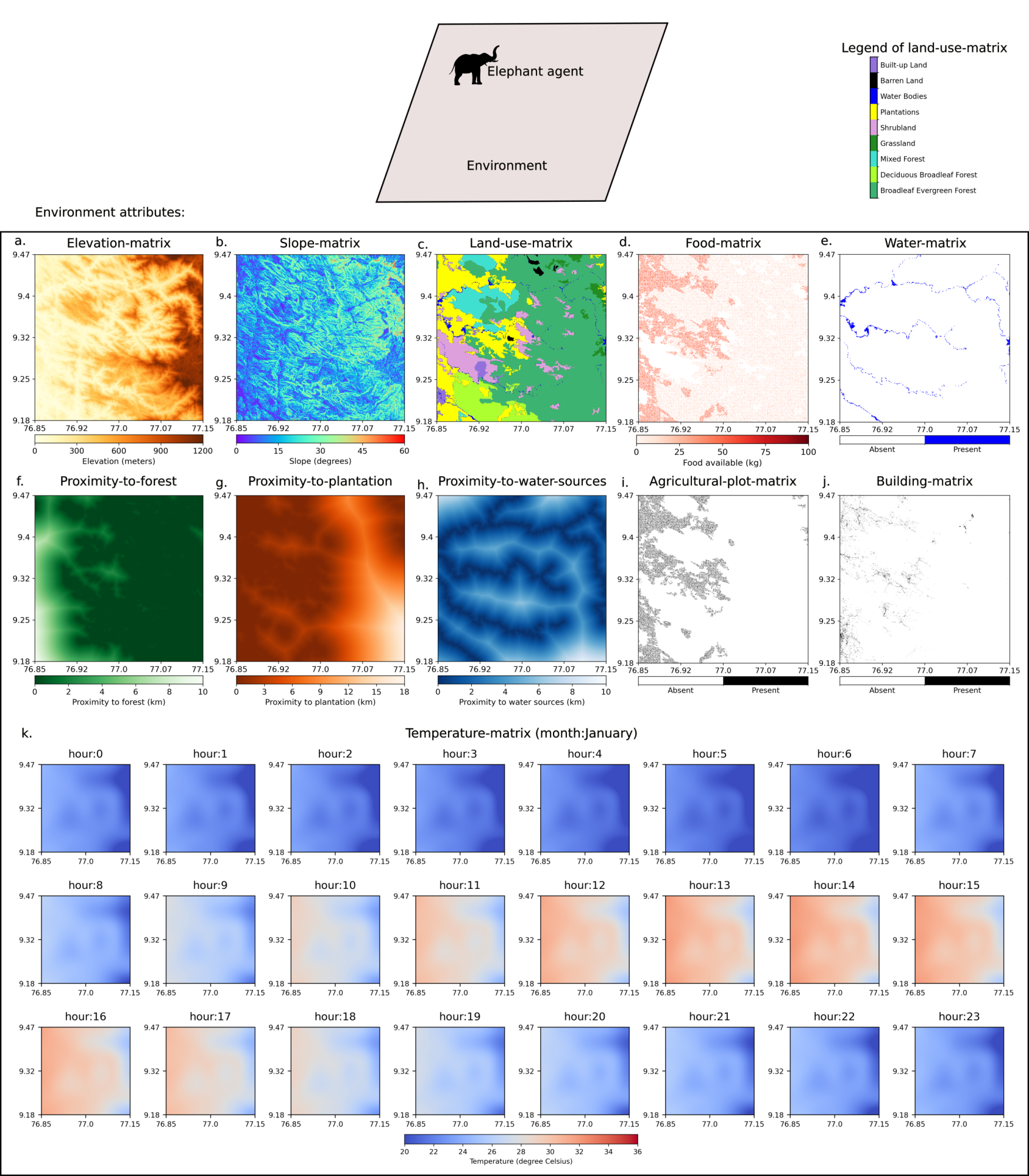}
\caption{\textit{The attributes of the environment used within the model.} The dynamic attributes include the food matrix (panel d) and the temperature matrix (panels k). The food matrix is updated each time the elephant agent consumes food from the landscape, while the temperature matrix is updated hourly, corresponding to the simulation month. All other environmental attributes (panels a-c,e-j) are static and remain unchanged throughout the simulation. Details of each attribute in Sect.~\ref{State variables}.}
\label{fig:simulation-grid-layers}
\end{figure*}

\textbf{Environment} The ABM is designed to be spatially explicit so that space use characteristics and emergent conflict patterns could be simulated. The environment is modelled as two-dimensional layers, discretized into landscape cells. The movement decision of elephants in each landscape cell is influenced by the landscape characteristics captured by the following attributes (Fig. \ref{fig:simulation-grid-layers}). 


\begin{enumerate}
    \item \textit{Elevation-matrix}: This attribute captures the elevation value above sea level in meters in each landscape cell of the simulation domain. It is obtained from a Digital Elevation Model (DEM) of the simulation domain. 

    \item \textit{Slope-matrix}: This attribute captures the gradient of each landscape cell in the simulation domain. Since the movement decisions of the elephant agents are dictated by the slope of the landscape, a DEM is used to infer the gradient and make appropriate movement choices.

    \item \textit{Land-use-matrix}: This attribute captures the land use class of each landscape cell of the simulation domain. It is obtained from the Land Use Land Cover (LULC) map. This information is used by different sub-models of the ABM to represent the heterogeneity within the landscape. There are 19 different classifications for land use: Deciduous Broadleaf Forest, Cropland, Built-up Land, Mixed Forest, Shrubland, Barren Land, Fallow Land, Wasteland, Water Bodies, Plantations, Aquaculture, Mangrove Forest, Salt Pan, Grassland, Evergreen Broadleaf Forest, Deciduous Needleleaf Forest, Permanent Wetlands, Snow and Ice, and Evergreen Needleleaf Forest. Of these, the simulation domain has only 9 land use classes (Fig.~\ref{fig:map_detailed} (b)). 
    
    \item \textit{Food-matrix}: This attribute captures the amount of food available to the elephant agents in each landscape cell of the simulation domain. This information is used by elephant agents to make foraging and movement choices to satisfy their dietary requirements. The \textit{food-matrix} is updated if the elephant agent consumes food while present in that cell. 
    
    
   
    \item \textit{Water-matrix}: This attribute captures the amount of water available in each landscape cell of the simulation domain. This information is used by elephant agents to satisfy their water needs and thermoregulation requirements. Each cell is a binary value that indicates whether or not a water source is present in the cell. Unlike the \textit{food-matrix}, it is assumed that the quantity of water does not decrease with its consumption by an elephant agent in the present prototype ABM.

    \item \textit{Proximity-maps}: This attribute captures how close each landscape cell is to a particular category of land use. Specifically, three different proximity maps are used: proximity to plantations, forests (Deciduous Broadleaf Forest, Mixed Forest and Broadleaf Evergreen Forest), and water sources. The magnitude of each cell of the matrix represents the proximity to that type of land use. These three maps are used in the decision-making process of the elephant agent's cognition model.

    \item \textit{Agricultural-plot-matrix}: This attribute captures the presence or absence of agricultural plots in the landscape cells. This information is used with a probability of damage to assess conflict scenarios.

    \item \textit{Building-matrix}: This attribute captures the presence or absence of buildings in the landscape cell. This information is used with a probability of damage to assess conflict scenarios.

    \item \textit{Temperature-matrix}: This attribute captures the spatio-temporal changes in the temperature values of each landscape cell. This information is a crucial component in the decision-making process of our model and is utilized in multiple submodels of the elephant agent's behavior and environmental interactions.
\end{enumerate}
\paragraph{State variables}\mbox{}\\ \label{State variables} 


Table~\ref{tab:elephant-agent's-state-variables} lists the 29 state variables of the elephant agent, and Table~\ref{tab:landscape-cell-attributes} lists the 9 state variables of the environment used to characterize the entities in the model. The state variables \textit{fitness} and \textit{aggression}, and the behavioral state \textit{thermoregulation} are defined below. The other state variables are self-explanatory from the name.

The \textit{fitness} of the elephant agent is directly related to its energy stores and its ability to survive. In the simulation, the agent's \textit{fitness} is measured on a scale of 0 to 1. A \textit{fitness} value of 0 indicates death, while a value of 1 indicates maximum health. 

The \textit{aggression} level of the elephant agent serves as an indicator of its willingness to take risks and enter plantations to raid crops. This metric measures the probability that the agent chooses riskier cells near plantations on a scale of 0 to 1, either when there is a shortage of food in the forest or when it has adapted to crops, and there is an abundant food supply in croplands. 

\textit{Thermoregulation in elephants}: Mammals like elephants have less surface area available for heat dissipation due to their larger body size. Therefore, thermal stress and thermoregulation play a very important role in elephant ecology, especially in areas of high ambient temperatures \citep{kinahan2007,weissenbock2012,thaker2019}. Thermoregulation refers to the requirement of elephants to maintain their core body temperature at a constant level at all environmental temperatures. In our ABM, thermoregulation actions of shade seeking/resting, wallowing and moving towards or visiting a water source are considered.

\begin{table*}[h!]
\caption{The elephant agent's state variables}\label{tab:elephant-agent's-state-variables}
\resizebox{0.8\textwidth}{!}{
    \begin{tabular}{p{0.15\textwidth}p{0.65\textwidth}p{0.05\textwidth}p{0.12\textwidth}p{0.12\textwidth}}
    \toprule
    \textbf{Parameter} & \textbf{Description} & \textbf{Type} & \textbf{Range} & \textbf{Unit} \\ 
    \midrule
    \textit{unique-ID} & Unique identifier of the agent, special name containing numeric and non-numeric characters (static) & string & & \\
    \midrule
    \textit{age} & Age of the elephant (dynamic) & integer & 15 to 60 & Years \\
    \midrule
    \textit{body-weight} & Body weight of the elephant (dynamic) & float & 3250 to 4000 & kg \\
    \midrule
    \textit{daily-dry-matter-intake} & Daily dry matter dietary requirement of the elephant (dynamic) & float & 48.75 - 76 & kg \\
    \midrule
    \textit{knowledge-from-fringe} & The distance from the forest fringe where the elephant knows food availability (static) & float & 0 - undefined & meter \\
    \midrule
    \textit{percent-memory-elephant} & The percentage of landscape cells known to the elephant in terms of food and water availability at the start of the simulation (static) & float & 0 - 100 & percentage \\
    \midrule
    \textit{(current-x, current-y)} & Current location of the elephant (dynamic) & tuple of floats &  Extent of the simulation area & agent location in WGS84$^{*}$ \\
    \midrule
    \textit{(next-x, next-y)} & Location to which the elephant moves next (dynamic) & tuple of floats &  Extent of the simulation area & agent location in WGS84$^{*}$ \\
    \midrule
    \textit{mode} & Current mode or behavioral state of the elephant (dynamic) & string & \textit{random-walk} / \textit{foraging} / \textit{thermoregulation} / \textit{escape-mode} & \\
    \midrule
    $p_t$ & Thermoregulation probability (dynamic) & float & 0 - 1 & \\
    \midrule
    \textit{thermoregulation-threshold} & Temperature value above which the elephant engages in \textit{thermoregulation} mode (static) & float & 0 - 1 & \\
    \midrule
    \textit{radius-food-search} & Radius within which the elephant searches for food in its \textit{memory-matrix} (static) & float & 0 - undefined & meter \\
    \midrule
    \textit{radius-water-search} & Radius within which the elephant searches for water in its \textit{memory-matrix} (static) & float & 0 - undefined & meter \\
    \midrule
    \textit{radius-forest-search} & Radius within which agent remembers the forest boundary, to escape in case of conflict with humans (static) & float & 0 - undefined & meter \\
    \midrule
    \textit{fitness} & Fitness of the elephant (dynamic) & float & 0 - 1 & \\
    \midrule
    \textit{aggression} & Aggression of the elephant (static) & float & 0 - 1 & \\
    \midrule
    \textit{movement-fitness-deprecation} & Movement cost; the factor by which \textit{fitness} decreases at each time step (static) & float & 0 - 1 & \\
    \midrule
    \textit{fitness-increment-when-eats-food} & Factor by which the \textit{fitness} value increment when the agent consumes food (static) & float & 0 - 1 & \\
    \midrule
    \textit{fitness-increment-when-thermoregulates} & Factor by which the \textit{fitness} value increment when the agent thermoregulates when in \textit{thermoregulation} mode (static) & float & 0 - 1 & \\
    \midrule
    \textit{fitness-threshold} & \textit{Fitness} value below which the elephant only engages in \textit{foraging} mode (static) & float & 0 - 1 & \\
    \midrule
    \textit{terrain-radius} & Parameter in terrain cost function (static) & float & 0 - undefined & meter \\
    \midrule
    \textit{tolerance} & Parameter in terrain cost function (static) & float & 0 - undefined & \\
    \midrule
    \textit{num-days-water-source-visit} & Number of days elapsed since last visited a water source (dynamic) & integer & 0 - undefined & Days \\
    \midrule
    \textit{num-days-food-deprivation} & Number of days elapsed since the daily dietary requirement was satisfied (dynamic) & integer & 0 - undefined & Days \\
    \midrule
    \textit{prob-crop-damage} & Probability with which the agent damages crop when entered an agricultural cell (static) & float & 0 - 1 & \\
    \midrule
    \textit{prob-infrastructure-damage} & Probability with which the agent damages infrastructure when entering a building cell (static) & float & 0 - 1 & \\
    \midrule
    \textit{danger-to-life} & If the elephant perceives a danger to its life (dynamic) & boolean & True/False &  \\
    \midrule
    \textit{disturbance-tolerance} & If the elephant agent is used to foraging in the presence of humans (static) & boolean & True/False &  \\
    \midrule
    \textit{food-habituation} & If the elephant agent prefers to consume agricultural crops (static) & boolean & True/False &  \\
    \bottomrule
    \end{tabular}
}
\flushleft
\scriptsize{[*] WGS84: World Geodetic System 1984 (EPSG:3857, Pseudo-Mercator)}
\end{table*}
\begin{table*}[h!]
\caption{The attributes of the landscape cells}
\label{tab:landscape-cell-attributes}
\resizebox{0.985\textwidth}{!}{
\begin{tabular}{p{0.15\textwidth}p{0.55\textwidth}p{0.15\textwidth}p{0.12\textwidth}p{0.12\textwidth}}
\toprule
\textbf{Parameter} & \textbf{Description} & \textbf{Type} & \textbf{Range} & \textbf{Unit} \\ 
\midrule
\textit{Elevation-matrix} &  Digital Elevation Model of the study area (static) & array of floats & 0 - undefined & meter \\
\midrule
\textit{Slope-matrix} &  Slope map of the study area (static) & array of floats & 0 - undefined & degrees \\
\midrule
\textit{Land-use-matrix} & Land Use Land Cover map of the study area (static) & array of integers & 0 - 18 & categorical \\
\midrule
\textit{Food-matrix} & Food availability matrix within the landscape (dynamic)  & array of floats & 0 - 100 & kilograms \\
\midrule
\textit{Water-matrix} & Water availability matrix within the landscape (static) & array of integers & 0 - 1 & categorical \\
\midrule
\textit{Proximity-maps} & Proximity of the current cell to a particular category of land use cell (static) & array of floats & 0 - undefined & meter \\
\midrule
\textit{Agricultural-plot-matrix} & Presence/absence of the agricultural plot within a landscape cell (static) & array of integers & 0 - 1 & categorical \\
\midrule
\textit{Building-matrix} & Presence/absence of the buildings within a landscape cell (static) & array of integers & 0 - 1 & categorical \\
\midrule
\textit{Temperature-matrix} & Hourly temperature within the study area (dynamic) & array of floats & 0 - undefined & degree Celsius\\
\bottomrule
\end{tabular}
}
\end{table*}
\paragraph{Scales} 
\begin{enumerate}
    \item \textbf{Spatial scale}: The simulation was carried out on a landscape area of approximately 1100~$km^2$, discretized into cells of $30\times 30~m$. This spatial scale was chosen to accurately capture the scales at which elephants might make movement decisions \citep{Diaz2021}. Thus, $1069\times 1070$ spatial cells are in our ABM.

    \item \textbf{Temporal scale}: The temporal scale of the simulation was set at 5 minutes to match the spatial scale choice. This fine resolution also helps to capture interactions with human agents in the future (not included in the current prototype ABM). Furthermore, movement data (Sect.~\ref{sec:field_survey_and_data_collection}) were available at this resolution to calibrate our movement model.
\end{enumerate}
\subsection{Process overview and scheduling}

Process scheduling was reached through an iterative modeling process that involved expert opinion, an extensive literature survey including academic publications and newspaper articles, and the rationality and understanding of the modelers \cite{Oommen_2017_a, Oommen_2017_b, Oommen_2018}. 


The drivers of the conflict could be many. We attempted to build a model that accommodates the multiple drivers of conflict and tested the model against the various patterns expected under the given ecological and psychological conditions. The most challenging part of model development was approximating the cognition of the elephant agents and determining the most important processes within its decision-making scheme. The elephant agents' process scheduling was finalized to be of the following order after a thorough study involving conceptual as well as experimental analysis.
    
\begin{enumerate}
    
    \item Set the food goal for the agent according to its body weight at the beginning of the day. If the previous day's food goal was not satisfied or the elephant did not visit a water source, update its fitness accordingly. 
    
    \item If the elephant agent senses a danger to its life, handling this situation becomes the priority (\textit{escape-mode}. If there is no danger to life, then the elephant agent proceeds with its activity (\textit{foraging}, \textit{thermoregulation} or \textit{random-walk} mode).
    
    \item If the elephant agent moves into an agricultural-plot or building cell, it inflicts some damage based on the corresponding damage probabilities.


    
    \item The elephant agent consumes food and drinks water if it encounters a food or water cell.
        
\end{enumerate}
    
Furthermore, the four different activity modes (\textit{escape-mode}, \textit{thermoregulation}, \textit{foraging} and \textit{random-walk}) and the corresponding subprocesses were designed to replicate the patterns and observations reported in the literature, and that emerged through expert insights and field surveys. The proposed process is the best approximation that the modelers could arrive at keeping in mind the utility of the prototype ABM, model complexity, and computational cost.

\begin{figure*}[h!]
\centering
\includegraphics[width=0.5\linewidth]{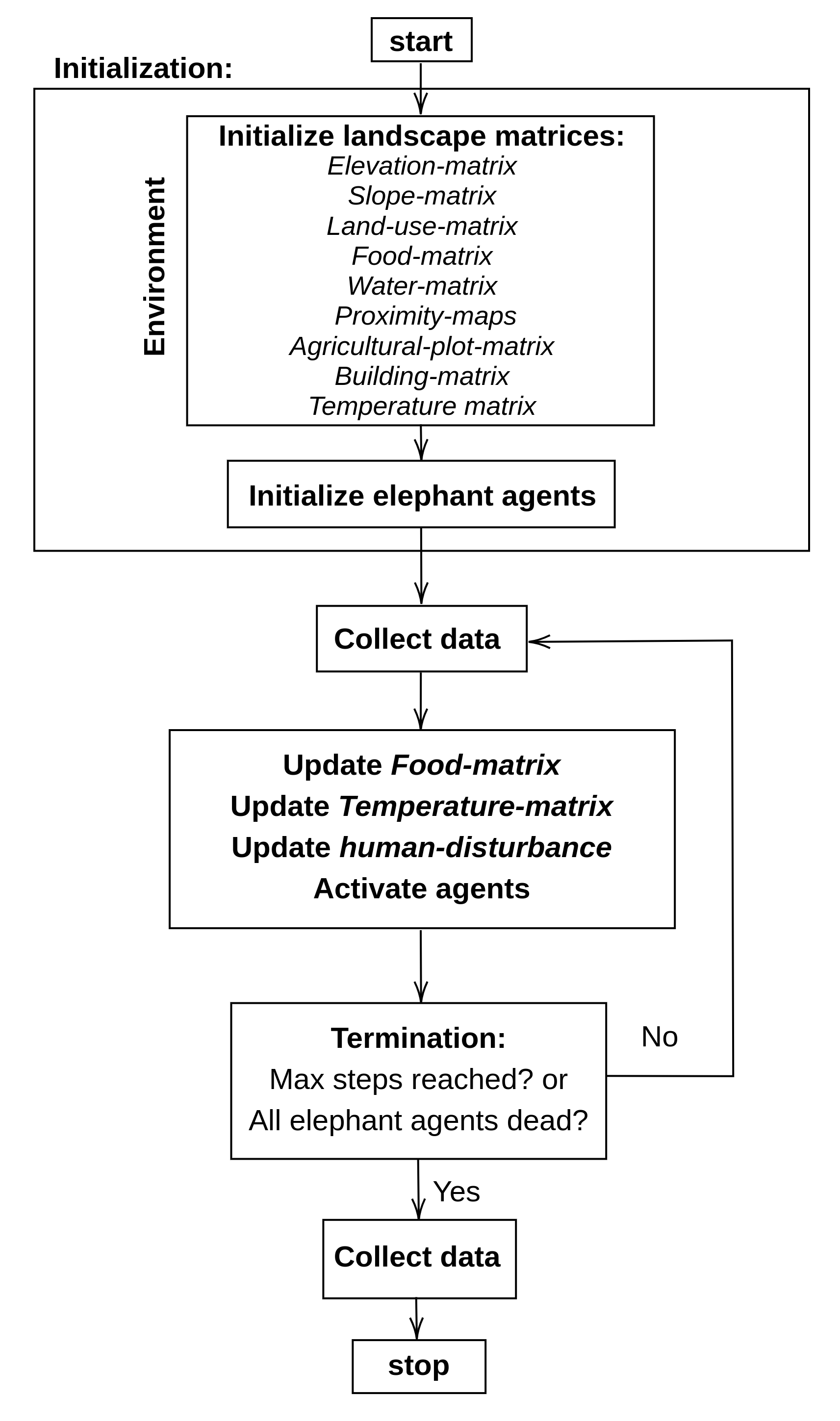}
\caption{A summary of the model schedule and the process updates.}
\label{fig:process-schedule}
\end{figure*}

\begin{figure*}[h!]
\centering
\includegraphics[width=0.75\linewidth]{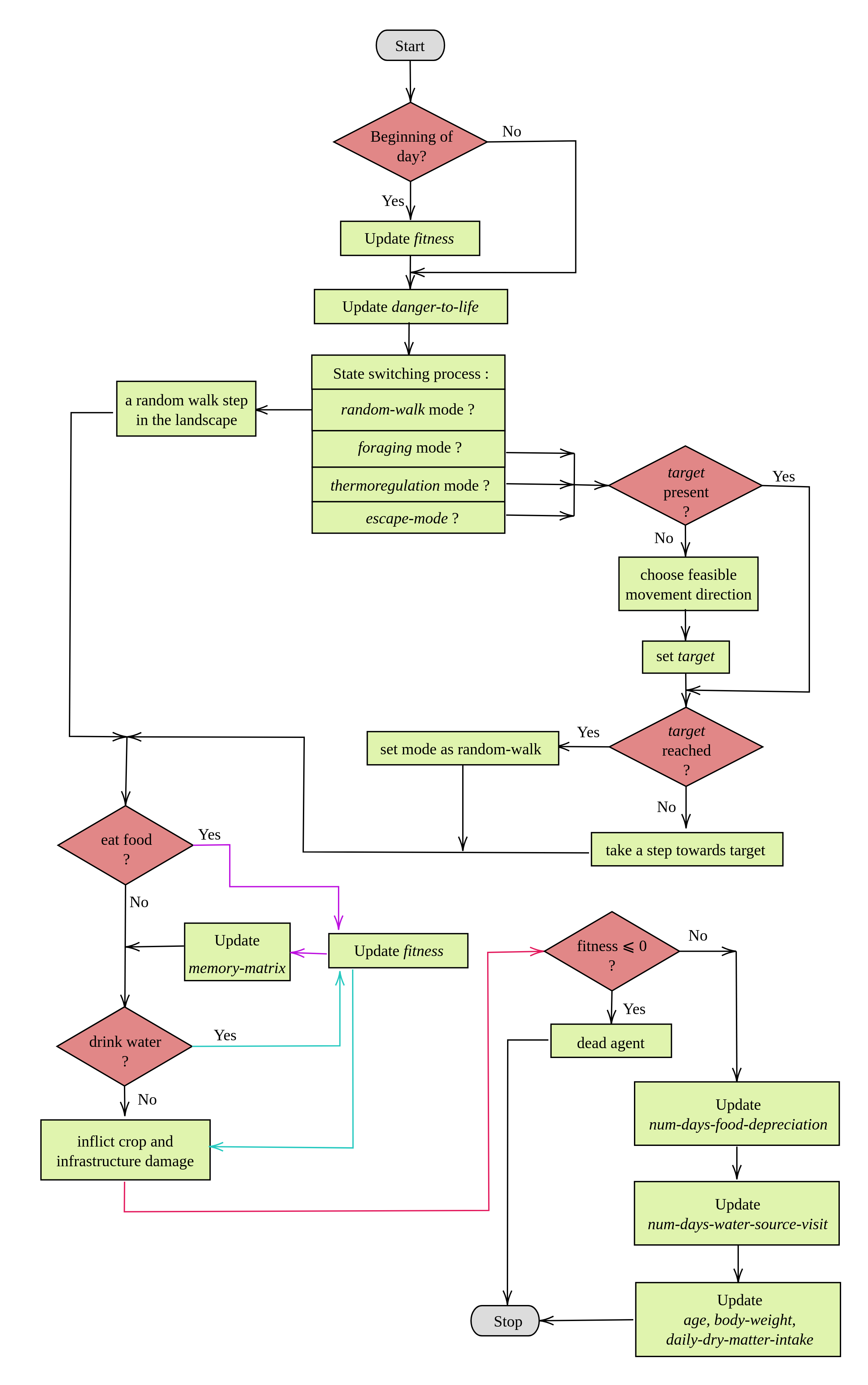}
\caption{The flow chart of the decision-making process of the elephant agent at each time step.}
\label{fig:elephant-decision-making}
\end{figure*}

 Table \ref{tab:process-overview-model} lists the 6 processes in the ABM, and Fig.~\ref{fig:process-schedule} shows the schedule of these processes. Of these, the `activate agents' process triggers the processes associated with the elephant agent. Table \ref{tab:process-overview-elephants} lists the 21 processes in the elephant cognition model, and Fig.~\ref{fig:elephant-decision-making} shows the scheduling of these processes.

\begin{table*}[h!]
\caption{The process overview of the model}
\label{tab:process-overview-model}
    \begin{tabular}{p{0.005\textwidth}p{0.25\textwidth}p{0.60\textwidth}}
    \toprule
    \textbf{} & \textbf{Process} & \textbf{Description} \\ 
    \midrule
    1. & Initialize environment & The process that initializes different landscape attributes \\
    2. & Initialize elephant agents & The process associated with initialization and distribution of the elephant agents within the landscape \\
    3. & Update \textit{Food-matrix} & The process that updates the landscape \textit{Food-matrix}  \\
    4. & Update \textit{Temperature-matrix} & The process that updates the landscape \textit{Temperature-matrix} \\
    5. & Update \textit{human-disturbance} & The process that updates human disturbance/human-activity within the landscape \\
    6. & Activate agents & The process that determines the order in which the agents are activated \\
    \bottomrule
    \end{tabular}
\end{table*}
\begin{table*}[h!]
\caption{The process overview of the elephant agents}
\label{tab:process-overview-elephants}
\begin{tabular}{p{0.010\textwidth}p{0.15\textwidth}p{0.725\textwidth}}
\toprule
\textbf{} & \textbf{Process} & \textbf{Description} \\ 
\midrule
1. & Movement model & Determines how the \textit{next-step} for the agent to move is sampled \\
2. & Update \textit{danger-to-life} & Updates whether the agent perceives a threat to its life \\
3. & Behavioral state switching & Chooses a behavioral mode based on the internal state of the agent as well as external factors. \\
4. & Feasible movement direction & Chooses a feasible direction for the agent to move depending on the movement cost in different directions \\
5. & \textit{random-walk} mode & Decision-making process when the agent chooses to walk randomly in the landscape \\
6. & \textit{foraging} mode & Decision-making process when the agent is in the foraging mode and moving toward a food target \\
7. & \textit{thermoregulation} mode & Decision-making process when the agent is in the \textit{thermoregulation} mode\\
8. & \textit{escape-mode} & Decision-making process of escaping to the forest in case of danger to life \\
9. & \textit{target} to eat food & Decision-making process of choosing a food source target \\
10. & \textit{target} to thermoregulate & Decision-making process of choosing a thermoregulation target \\
11. & \textit{target} for escape & Decision-making process of choosing a forest cell to escape to in case of danger to life \\
12. & Eat food & The process that determines how the agent consumes food from the landscape \\
13. & Update \textit{fitness} & Includes all mechanisms by which the fitness of the agent is updated \\
14. & Crop and infrastructure damage & Decision-making scheme of inflicting damage to crops and property \\
15. & Death & The process that determines if the agent is alive or dead \\
16. & Update \textit{age} & The process that updates the \textit{age} of the agent \\
17. & Update \textit{body-weight} & The process that updates the \textit{body-weight} of the agent \\
18. & Update \textit{daily-dry-matter-intake} & The process that updates the \textit{daily-dry-matter-intake} of the agent \\
19. & Update \textit{memory-matrix} & The process that updates the \textit{memory-matrix} of the agent\\
20. & Update \textit{num-days-water-source-visit} & The process that updates the \textit{num-days-water-source-visit} of the agent \\
21. & Update \textit{num-days-food-deprivation} & The process that updates the \textit{num-days-food-deprivation} of the agent \\
\bottomrule
\end{tabular}
\end{table*}
\subsection{Design concepts}\mbox{}

\paragraph{Basic principles}\mbox{}\\
The following are the general concepts used in model design:

\begin{enumerate}

    \item The elephant agent's movement pattern and space use reflect the mountainous terrain of the study area. Specifically, slope and natural barriers that restrict movement are taken into account. The movement is either in an \textit{exploratory} or \textit{encamped} state as reported in the literature.
    
    \item The optimal foraging theory is used to satisfy the daily dietary requirements and corresponding movement choices. Changes in elephant behavior with the changing external environment (food abundance and scarcity scenarios) are also considered.

     \item Elephants thermoregulate when the ambient temperature increases.

     \item The effect of human disturbance on the elephant agent's movement and space use is considered.

     \item The spatial memory attributes of elephants reported in the literature are captured in their decision-making.
    
\end{enumerate}

To achieve the above basic principles, the model relies on the following underlying assumptions and conceptual theories. 

\textbf{Movement modeling of elephants}: Several studies have shown two distinct states, viz. \textit{exploratory} and \textit{encamped} in elephant relocation data. Elephants have been observed to exhibit increased exploratory behavior in corridors and increased encamped movements near rivers during the dry season \citep{Vogel2020}. A study of the turning angles of elephant paths revealed two distinct movement states, \textit{(i)} high tortuosity paths with frequent turns and \textit{(ii)} straight line paths with minimal turning \citep{Bailey2021}. Thus, two-state Hidden Markov Models (HMMs) with \textit{exploratory} and \textit{encamped} states have emerged as an appropriate choice for modeling elephant movement \citep{Beirne2021}. This model is used in our ABM.


\textbf{Thermoregulation in elephants}: Elephants, as large mammals, have a limited surface area for heat dissipation, making thermoregulation crucial for their survival, especially in hot environments \citep{kinahan2007, weissenbock2012, thaker2019}. Studies have shown that elephants thermoregulate by adjusting their movement choices based on ambient temperature and landscape characteristics \citep{kinahan2007}. The core body temperature of elephants fluctuates throughout the day, gradually increasing until late in the evening and decreasing at night \citep{kinahanBodyTemperatureDaily2007}. Their behavior also adapts to heat stress, with elephants seeking shade, increasing water-related activities (e.g., wetting, wallowing and bathing), and resting more \citep{mole2016}. The availability of resources also influences the movement of elephants and the use of space \citep{yolandi2008, jachowski2013}. Seasonal changes in vegetation productivity affect its spatial distribution, and elephants depend more on water sources during dry seasons \citep{young2009}. 

The availability of water is important for thermoregulation and significantly affects the movement of Asian and African elephants. For example, African elephants in Kruger National Park exhibit shorter and faster movements toward water during warmer periods \citep{thaker2019}. In areas with limited water, such as the Tsavo Conservation Area, elephants showed directional movement toward water sources, females rarely strayed far, and males traveled longer distances during dry seasons \citep{Wato2018}. Feeding behavior also adapts to ambient temperature, with Asian elephants showing bimodal peaks in activity during cooler mornings and evenings and resting more during the hot midday period \citep{Baskaran2010}. These studies highlight the crucial role of thermoregulation in elephant decision-making. During the wet months, elephants do not need to visit water sources explicitly for drinking (Sanjeeta Sharma Pokharel, personal communication, April 19, 2024), as rainfall creates multiple accessible surface water sources. In the dry months, these sources deplete, leading elephants to stay near perennial water sources like rivers and streams. Visiting water sources for drinking water and participating in other activities such as wetting, wallowing, and bathing are the aspects of thermoregulation included in our ABM.


\textbf{Effect of human disturbance on elephants}: Several studies have documented the effects of human disturbance on elephants, highlighting its influence on their physiology, activity patterns, and behavior. Elephants exposed to human activity exhibited elevated cortisol and estradiol levels, indicating stress \citep{Tang2020}. It was observed that elephant activity was lower in high human disturbance areas, and elephants preferred nocturnal movement to avoid humans \citep{Wilson2021}. In the Western Ghats, human interference, including resource competition, habitat degradation, and hunting, negatively affected elephant presence \citep{Jathanna2015}. This factor also impacts their spatial distribution, movement patterns, vigilance, foraging, and social interactions \citep{Barnes1991,Kumar2010,Srinivasaiah2012,Gaynor2018}. The questionnaire-based survey in the study area insights revealed that sounds deterred elephants from crop lands. Considering the significant effects of human disturbance on elephants, it is included in our ABM.

\textbf{\textit{Memory-matrix} of elephants}: Elephants use their exceptional olfactory senses to identify water sources \citep{Wato2018,wood2023} and food availability \citep{schmitt_2018, plotnik_2019, nevo_2020}. Their excellent memory allows them to retain landscape and food knowledge over time. Our ABM includes submodels to simulate realistic foraging patterns leveraging sensory perception and spatial memory. We use a \textit{memory-matrix} to capture the elephant's memory of these sources. Complete knowledge of water is assumed, while a \textit{percent-memory-elephant} state variable changes adaptively during foraging to represent the percentage of landscape cells with known food. A nuance is that elephant memory varies with age/sex and factors like deforestation. In this ABM, we simulate a 40-year-old male elephant over one year, making a simple \textit{memory-matrix} sufficient. 

\paragraph{Emergence}


The emergent space use pattern of the elephant agents, and the emergent crop raid and infrastructure damage incidents were investigated.
\paragraph{Fitness}

The elephant agent tries to maximize its \textit{fitness} objective while satisfying \textit{thermoregulation} needs. In our ABM, \textit{fitness} is defined as a state variable that tracks the energy level and the ability of the elephant agent to survive. The \textit{fitness} is updated explicitly at each timestep. 
\paragraph{Adaptation}

The elephant agent's movement decision and behavioral state switching have been modeled to adapt to changing internal state and external environment via empirical rules and probabilistic models. The goal of adaptation is to improve \textit{fitness} through direct and indirect \textit{fitness} maximization objectives.

\paragraph{Objectives}


The primary objective of the elephant agent is to meet its daily nutritional and thermoregulation requirements while taking into account the cost of movement across different terrains. Elephant agent's behavioural state switching and movement has been modelled as adaptive of its internal state as well as response to the external environment, so as to improve its \textit{fitness}. Both direct and indirect fitness maximization have been included in the model. All of these adaptive behaviours are modelled via simple empirical rules and probabilistic models.
\paragraph{Learning, Prediction}


In the present model, learning agents and prediction of future conditions were not implemented.


\paragraph{Sensing}


The elephant agents can sense the elevation of the surrounding landscape and know the food and water availability within the simulation area to some extent through their memory matrices, the land use classification types, and the spatio-temporal temperature changes within the simulation domain. Elephant agents can also assess the level of human activities on the forest fringe to make choices about the temporal aspects of crop raiding depending on their risk profile.
\paragraph{Interaction}

Currently, only one bull elephant agent is considered. Thus, there are no interactions between multiple agents.
\paragraph{Stochasticity}
Submodels for elephant movement, choice of feasible movement direction, behavioral state-switching, target to eat food, thermoregulate and escape, and food consumption are all partially modeled as stochastic processes (Sect.~\ref{section:sub-models}). The input environmental variables are not considered stochastic and their climatological values are used in the simulation.

\paragraph{Collectives}
There are no collectives in this model.
\paragraph{Observation}

The following data are collected at each time step of the ABM simulation: crop damage incidents, infrastructure damage incidents, number of elephant deaths, location (\textit{current-x, current-y}), \textit{fitness}, \textit{mode}, \textit{num-days-water-source-visit}, and \textit{num-days-food-deprivation} of the elephant agents.
\subsection{Initialization}
\label{section:initialization}

All ABM simulations start with the same initial location for the bull elephant. This location is in a carefully chosen Broadleaf Evergreen Forest cell, 7~$km$ from the croplands and 600~$m$ above sea level. The selection of this point was made because a spatial distribution study in a comparable location \citep{Surendra2008} found that elephant dung counts were predominantly recorded in evergreen forests in altitude ranges of 600 and 1200~$m$, with a presence within 300 to 1300~$m$.

In all experiments, 40-year-old elephant agents with a body weight of 4000~$kg$ were used. This elephant agent must consume 68~$kg$ of food per day to meet its daily dietary requirement \cite{Ullrey1997}. The initial behavioral state of the elephant agent was randomly chosen from one of the \textit{random-walk} and \textit{foraging} modes. The elephant agents were assumed to have met their dietary requirement (\textit{num-days-food-deprivation} = 0) and visited a water source (\textit{num-days-water-source-visit} = 0) at the beginning of the simulation. The \textit{fitness} of all agents is initialized to their maximum value of 1. The state variable \textit{aggression} was initialized according to the experiment conducted. The value of the \textit{knowledge-from-fringe} variable was set to 1500~$m$, and the \textit{memory-matrix} of the elephant agent was initialized accordingly. This decision was made based on the following reasoning. It was assumed that the elephant agent had an approximate understanding of the type of food crops and the availability of food within the cells of its \textit{memory-matrix}. The patterns of crop raid incidents vary depending on the environmental and behavioral characteristics of the elephant groups involved, as observed at different locations. For example, in a study conducted in a similar landscape in South India, crop raiding incidents were most common within 2~$km$ from the forest boundary, with occasional incidents even reported beyond 5~$km$ \citep{Kalyanasundaram2013}. A study of conflict interactions involving the African elephant found that crop raid incidents typically occurred within an average distance of 1.54~$km$ from daytime elephant refuges \citep{Max2010}. However, there have also been reports of incidents in which the extent of damage per farm increased with increasing distance from the boundary of the protected area \citep{Abraham2022}. The questionnaire data collected from the field recorded conflict reports at a distance of 1500~$m$ from the fringe. Therefore, the \textit{knowledge-from-fringe} variable was set to this value.

The remaining state variables were set to their initial values through a meticulous calibration process using movement data (Sect.~\ref{sec:Calibration_of_the_model}).
\subsection{Input}
\label{section:input}

The input data are obtained from various external sources, either as is or slightly adjusted to fit the required format for accurate attribute representation.

The \textit{elevation-matrix} and \textit{slope-matrix} were created using DEM data from the SRTM repository \citep{SRTM_data}, available at a resolution of 30~$m$.

The \textit{Land-use-matrix} was created from LULC data for India (for the year 2005) from Landsat at 100~$m$ resolution \citep{LULC_data}, downscaled to 30~$m$ using QGIS \citep{QGIS_software}.

The \textit{water-matrix} was derived from downscaled LULC data at 30~$m$ resolution. If water is present in a landscape cell, the corresponding entry in the \textit{water-matrix} is 1, else 0. We assumed that the main source of water for the elephant agents is rivers and streams. Other sources such as waterholes and ponds are excluded in this study and will be considered in the future. 

The evergreen broadleaf, deciduous broadleaf, and mixed forests were grouped as forest land. The \textit{proximity-maps} for plantations, forests, and water sources were derived from the downscaled LULC using QGIS.

Agricultural land use data were collected through a questionnaire (Sect.~\ref{sec:field_survey_and_data_collection}), identifying seven categories: home gardens with varying percentages of rubber trees (0\%, 0-25\%, 25-50\%, 50-75\%, 75-100\%, 100\%), and a None category. Data on fruiting crops such as plantains, jackfruits, and mango were also collected. These data were used to initialize the \textit{agricultural-plot-matrix} by extrapolating statistics to the entire cropland within the simulation limits. Since elephants prefer fruiting crops in home gardens, these gardens were designated as agricultural plots. The fully grown rubber trees in these gardens were assumed to be less susceptible to elephant damage due to their height.

The \textit{building-matrix} was constructed from a population density map at 30~$m$ resolution obtained from Facebook's Data for Good program \citep{population_data}. Since this map was created by an image analysis of buildings, we populated the \textit{building-matrix} by identifying cells that had some population and designating it as a cell with a building.

The spatio-temporal temperature data was collected from the WorldClim data website [CRU-TS 4.06 \citep{Harris2020}]. Additional information on the processing and its application within the model is provided in Sect.~\ref{sec:calibrating_using_literature}.

The \textit{food-matrix} was initialized for different experimental setups and is detailed in Sect.~\ref{experimental_setup}.

\subsection{Sub-models}
\label{section:sub-models}
The sub-models associated with the 6 processes in Table~\ref{tab:process-overview-model} are as follows.
\paragraph{Initialize environment}\mbox{}\\
This process sets up the attributes of the environment state by considering the day, month, and year specified at the beginning of the simulation.
\paragraph{Initialize elephant agents}\mbox{}\\
The elephant agent is initialized as described in Sect.~\ref{section:initialization}.
\paragraph{Update \textit{food-matrix}}\mbox{}\\
The \textit{food-matrix} is updated whenever the elephant agent consumes food. The corresponding food value is decremented from the landscape cell in the \textit{food-matrix}. 
\paragraph{Update \textit{temperature-matrix}}\mbox{}\\
The \textit{temperature-matrix} is updated every hour of simulation time using the temperature data.
\paragraph{Update \textit{human-disturbance}}\mbox{}\\
To model nocturnal crop raiding, we used a time-dependent human disturbance factor. This factor is set such that it exceeded the elephants' 	\textit{disturbance-tolerance} from 7 am to 7 pm, simulating daytime human activity, and was set at a low value from 7 pm to 7 am, reflecting reduced nighttime human activity. This approach captured the impact of human activity on the movement of elephants during crop raids.


\paragraph{Activate agents}\mbox{}\\
This process activates the agents in the ABM and triggers the corresponding decision-making process of the agents at each time step, depending on its state variables and the attributes of the environment. 
When the elephant agent is activated, the following sub-models associated with the 21 processes in Table~\ref{tab:process-overview-elephants} are used.
\paragraph{Movement model}\mbox{}\\
At every discrete time step, the movement model takes the elephant agent from the \textit{current-x, current-y} to the \textit{next-x, next-y} using a \textit{step-length} and \textit{turning-angle}. Two distinct movement patterns are considered in the ABM: \textit{(i)} \textit{exploratory} movement towards a destination and \textit{(ii)} \textit{encamped} movement within the landscape. \textit{Exploratory} movement is characterized by longer \textit{step length}s with fewer turns. On the other hand, \textit{encamped} movement is characterized by shorter \textit{step-length}s and more turns. An HMM is used to define the movement models and is calibrated with relocation data (Sect.~\ref{sec:calibrating_HMM_parameters}). The \textit{step-length} is modeled as a Gamma random variable in both patterns. The \textit{turning-angle} is modeled as a von Mises random variable in the \textit{encamped} movement model, and as a uniform random variable in the \textit{exploratory} movement model.

The algorithm~\ref{alg:targeted_walk_step_selection} describes the procedure by which the elephant agents move in each discrete time step during a \textit{exploratory} walk movement. The target of this walk is either a food source, a water source, or a forest cell, determined by the state-switching process (Fig.~\ref{fig:elephant-decision-making}). Once the target cell is identified, the \textit{heading} of the agent is directed toward the target added with a uniform noise sampled from $U(-15^\circ,+15^\circ)$. The step lengths are sampled from the gamma distribution identified from the fit of the HMM model. 

The algorithm~\ref{alg:random_walk_step_selection} describes the procedure for the motion of the agent during \textit{encamped} movement. The \textit{step-length} and \textit{turning-angle} are sampled and motion is executed. 

\paragraph{Update \textit{danger-to-life}}\mbox{}\\
The elephant agent perceives danger to its life when \textit{human-disturbance} is greater than the agent's \textit{disturbance-tolerance} when it is in a plantation cell. In this condition, the agent's \textit{danger-to-life} state is set to True. 

\begin{algorithm}
    \caption{\textit{Exploratory} movement model}
    \begin{algorithmic}[1]
    \STATE $\textit{step-length} = gamma$ ($\mu=0.0398, \sigma=0.0378$)
    \STATE $\textit{heading} = arctan2(\textit{target} - (\textit{current-x}, \textit{current-y}))$
    \STATE $\textit{theta}$ = $U(-15^\circ,+15^\circ)$
    \STATE $\text{dx} = \textit{step-length} \times \sin (\text{\textit{heading} + \textit{theta}})$
    \STATE $\text{dy} = \textit{step-length} \times \cos (\textit{heading} + \textit{theta})$
    \STATE $\textit{next-x} = \text{dx} + \textit{current-x}$
    \STATE $\textit{next-y} = \text{dy} + \textit{current-y}$
    \end{algorithmic}
    \label{alg:targeted_walk_step_selection}
\end{algorithm}

\begin{algorithm}
    \caption{\textit{Encamped} movement model}
    \begin{algorithmic}[1]
    \STATE $\textit{step-length} = gamma$($\mu=0.0040, \sigma=0.0034$)
    \STATE $\textit{turning-angle} = vonMises$($\mu=-3.0232, \kappa=0.3336$)
    \STATE $\textit{heading} = \textit{prev-heading} + \textit{turning-angle}$
    \STATE $\text{dx} = \textit{step-length} \times \sin (\textit{heading})$
    \STATE $\text{dy} = \textit{step-length} \times \cos (\textit{heading})$
    \STATE $\textit{next-x} = \text{dx} + \textit{current-x}$
    \STATE $\textit{next-y} = \text{dy} + \textit{current-y}$
    \end{algorithmic}
    \label{alg:random_walk_step_selection}
\end{algorithm}
\paragraph{Behavioural state switching}\mbox{}\\
Algorithm \ref{alg:StateSwitching} describes the mode or behavioral state switching procedure. The elephant agent is modeled to be in one of four different behavioral states: \textit{random-walk}, \textit{foraging}, \textit{thermoregulation}, and \textit{escape-mode}. If the elephant agent perceives a danger to its life, it switches to \textit{escape-mode}. If there is no danger to its life, the elephant agent engages in either \textit{random-walk} or \textit{foraging} or \textit{thermoregulation} mode. The probability of state switching between the modes \textit{random-walk} ($p_{11}$) or \textit{foraging} ($p_{22}$) was determined from the HMM movement model (Table~\ref{tab:movement-parameters}). If the agent's \textit{fitness} is very low (less than a set \textit{fitness-threshold}), it only operates in \textit{foraging} mode as foraging requirements supersedes thermoregulation (Sanjeeta Sharma Pokharel, personal communication, April 19, 2024). The agent switches to \textit{thermoregulation} mode when the ambient landscape cell temperature rises above its \textit{thermoregulation-threshold}. If the agent is walking toward a target, it is said to have reached the target when the distance between its current location and the target location is less than half the spatial scale of the landscape cell. If the agent reaches its target location, it switches to \textit{random-walk} mode.

\begin{figure*}[h!]
    \centering
    \includegraphics[width=0.85\textwidth]{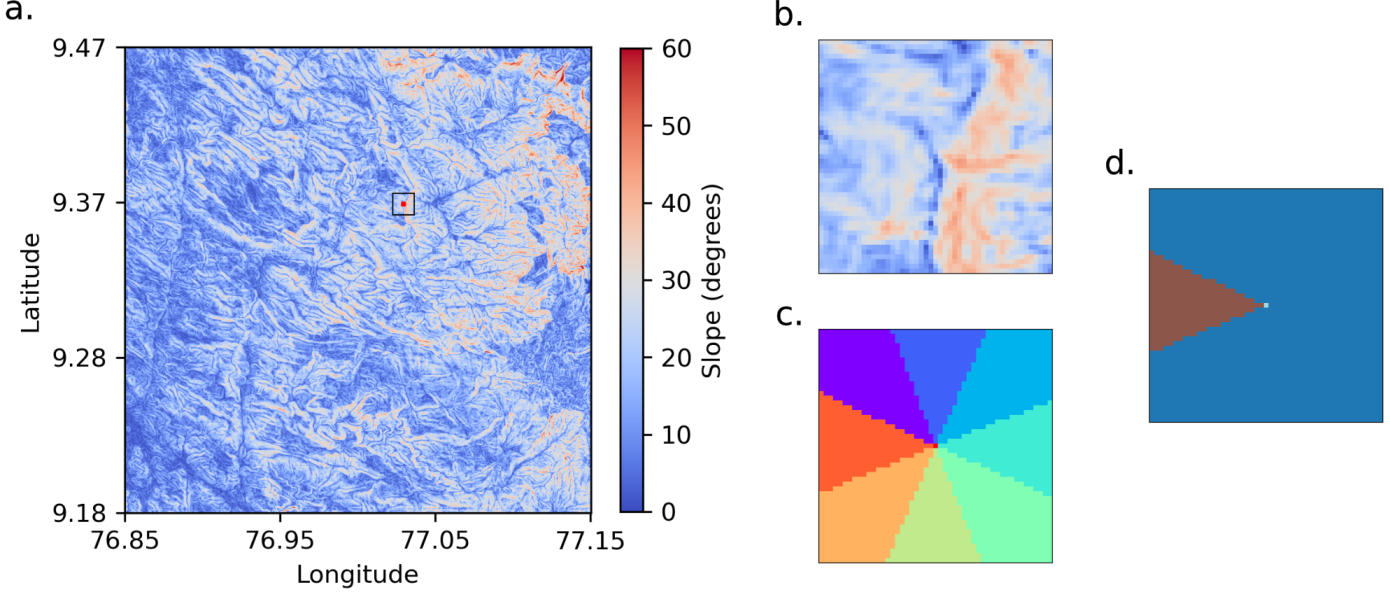}
    \caption{(a) The slope map of the simulation area with the red square highlighting the location of the elephant agent. The black square represents the \textit{terrain-radius} within which movement decisions are made. (b) The slope map within the \textit{terrain-radius} extent of the elephant agent (slope map within the black square in (a)). (c) The \textit{filter} to choose landscape cells along the cardinal and ordinal directions. (d) The feasible direction of movement selected according to the slope constraint is shown in brown.}
    \label{fig:direction_selection}
\end{figure*}

\begin{algorithm}[h!]
\caption{State switching}
\begin{algorithmic} [1]
\IF {\textit{danger-to-life} == True}
    \STATE mode = \textit{escape-mode}
\ELSE 
    \IF{\textit{ambient-temperature} $>$ \textit{thermoregulation-threshold}}
    \STATE mode = \textit{thermoregulation}
    \ELSIF{\textit{fitness} $<$ \textit{fitness-threshold}}
    \STATE mode = \textit{foraging}
    \ELSE 
    \STATE $\text{num} \sim Uniform$($0,1$)
        \IF{mode = \textit{random-walk}}
            \IF{$\text{num} < p_{11}$}
            \STATE mode = \textit{random-walk}
            \ELSE
            \STATE mode = \textit{foraging}
            \ENDIF
        \ELSIF{mode = \textit{foraging}}
            \IF{$\text{num} < p_{22}$}
            \STATE mode = \textit{foraging}
            \ELSE
            \STATE mode = \textit{random-walk}
            \ENDIF
        \ENDIF
    \ENDIF
\ENDIF
\end{algorithmic}
\label{alg:StateSwitching}
\end{algorithm}
\paragraph{Feasible movement direction}\mbox{}\\
We consider eight discrete directions, viz. north, south, east, west, north east, north west, south east, and south west for the movement of the elephant agent. However, in a given landscape cell, not all eight directions are feasible for movement, as the motion of the elephants is constrained by the gradient of the landscape. A movement cost is defined to help the agent select a feasible direction. First, a \textit{filter} matrix is used to select landscape cells in each of these directions (Fig. \ref{fig:direction_selection} (c)). Then, the movement cost is calculated as the sum of the slope values of the landscape cells that are greater than $30^\circ$ along these directions. A feasible direction is selected such that the movement cost is less than a set \textit{tolerance} along these directions. If there are several feasible directions, then one among these is chosen randomly with equal probability.
\paragraph{\textit{random-walk} mode}\mbox{}\\
In the \textit{random-walk} mode, the agent executes Algorithm~\ref{alg:random_walk_step_selection} to move to the \textit{next-step}. However, if the \textit{next-step} falls within a plantation cell, the agent chooses to move only if there is no danger to its life. Otherwise, it stays in the current cell and the simulation proceeds to the next time step.
\paragraph{\textit{thermoregulation} mode}\mbox{}\\
The agent selects a thermoregulation target within the feasible movement direction and subsequently moves towards this chosen target using the exploratory movement model. 
\paragraph{\textit{foraging} mode}\mbox{}\\
In \textit{foraging} mode, the agent moves to a food source using the \textit{exploratory} movement model. If the \textit{target} is a plantation cell, the \textit{human-disturbance} level must be less than the \textit{disturbance-tolerance} for activating exploratory walk. Otherwise, the \textit{escape mode} is activated.  
\paragraph{\textit{escape-mode}}\mbox{}\\
In escape mode, the agent chooses a target to escape and moves to this target using the \textit{exploratory} movement model.
\paragraph{\textit{target} to eat food}\mbox{}\\
The food target is selected from the agent's \textit{memory-matrix} within the \textit{radius-food-search} in the feasible movement direction. If the agent is not food-habituated or is not starved, then it randomly picks up a food target. On the other hand, if the agent has not been able to meet its dietary requirements for more than a \textit{threshold-num-days} (set as three in the present ABM) or if there is \textit{food-abundance} as measured by food in croplands more than 50\% the food in the forest and the agent is food-habituated, it chooses a \textit{target} closer to the croplands. Else, a random cell is selected from within the search radius in the feasible movement direction.
\paragraph{\textit{target} to thermoregulate}\mbox{}\\
In \textit{thermoregulation} mode, the agent selects the thermoregulation target as a landscape cell within the \textit{radius-forest-search} whose temperature is below the agent's \textit{thermoregulation-threshold}, or a water cell within the \textit{radius-water-search}. If no such landscape cell is found, the agent opts for a forest cell within the \textit{radius-forest-search} as the thermoregulation target. 
\paragraph{\textit{target} for escape}\mbox{}\\
The \textit{target} for escape is a forest cell, randomly picked from among the landscape cells within a distance of \textit{radius-forest-search} from the agent's current location. If there are no forest cells available within the search radius, the agent moves to a cell that is closer to the forest than its current cell within the search radius. To make this decision, the agent uses the landscape attribute \textit{proximity-to-forest}.  
\paragraph{Eat food}\mbox{}\\
The elephant agent consumes food if food is available within its current landscape cell. To be more realistic, it is assumed that the maximum food value in the landscape cell is not available to the elephant for consumption, a uniformly sampled food value between 0 and the maximum food value in the cell is available. When the food is consumed, the corresponding value of the food in the \textit{memory-matrix} and the landscape cell is decreased. The \textit{fitness} value of the agent increases with food consumption.
\paragraph{Update \textit{fitness}}\mbox{}\\
The fitness of the agent is updated in the following cases:
\begin{enumerate}
    \item When the elephant agent eats food, the \textit{fitness} increases by \textit{fitness-increment-when-eats-food}.
    \item When the elephant agent thermoregulates in \textit{thermoregulation} mode, the \textit{fitness} increases by \textit{fitness-increment-when-thermoregulates}.
    \item The \textit{fitness} decreases at each time step by \textit{movement-fitness-deprecation}.
\end{enumerate}
The \textit{movement-fitness-deprecation} is decreased from \textit{fitness} at each time step, while \textit{fitness-increment-when-eats-food} and \textit{fitness-increment-when-thermoregulates} is updated at the end of every day. 
\paragraph{Crop and infrastructure damage}\mbox{}\\
In case an elephant agent is within a cell with buildings or agricultural plots, damage is incurred using the corresponding damage probabilities, \textit{prob-infrastructure-damage} and \textit{prob-crop-damage}. 
\paragraph{Death}\mbox{}\\
The elephant agent is declared dead if its \textit{fitness} drops to zero.
\paragraph{Update \textit{age}}\mbox{}\\
The \textit{age} of the elephant agent is incremented by 1 year, every time the simulation completes a year. This is used to account for the growth of elephants.
\paragraph{Update \textit{body-weight}}\mbox{}\\
The \textit{body-weight} of the elephant agent is updated whenever its age is updated. The von Bertalanffy function is to update the body weight of the elephant agents corresponding to the increase in \textit{age} (Sect.~\ref{sec:calibrating_using_literature}).
\paragraph{Update \textit{daily-dry-matter-intake}}\mbox{}\\
The \textit{daily-dry-matter-intake} of the elephant agent is updated whenever its \textit{age} and \textit{body-weight} is updated. The \textit{daily-dry-matter-intake} is updated as 1.7\% of the \textit{body-weight} of the agent. 
\paragraph{Update \textit{memory-matrix}}\mbox{}\\
The \textit{memory-matrix} is updated whenever the elephant agent consumes food. The corresponding food value is decremented in memory. 
\paragraph{Update \textit{num-days-water-source-visit}}\mbox{}\\
The \textit{num-days-water-source-visit} is incremented by 1, for every day the agent is unable to visit a water source. If the agent visits a water source, \textit{num-days-water-source-visit} is set to zero.
\paragraph{Update \textit{num-days-food-deprivation}}\mbox{}\\
The \textit{num-days-food-deprivation} is incremented by 1, for every day the agent's food consumption was less than \textit{daily-dry-matter-intake}. If the agent has been able to satisfy its \textit{daily-dry-matter-intake}, then \textit{num-days-food-deprivation} is set to zero.
\subsection{Calibration of the model parameters and state variables}
\label{sec:Calibration_of_the_model}
The calibration of the model to set the movement model and other parameters of the submodel (Table~\ref{tab:movement-parameters} and Table~\ref{tab:model_calibration}), and the determination of the initial conditions of the agent's state variables were carried out using a five-step approach as follows.
\subsubsection{Identifying correlated random walk parameters by fitting a two-state Hidden Markov model to relocation data}
\label{sec:calibrating_HMM_parameters}

\begin{figure*}[h!]
    \centering
    \includegraphics[width=0.8\textwidth]{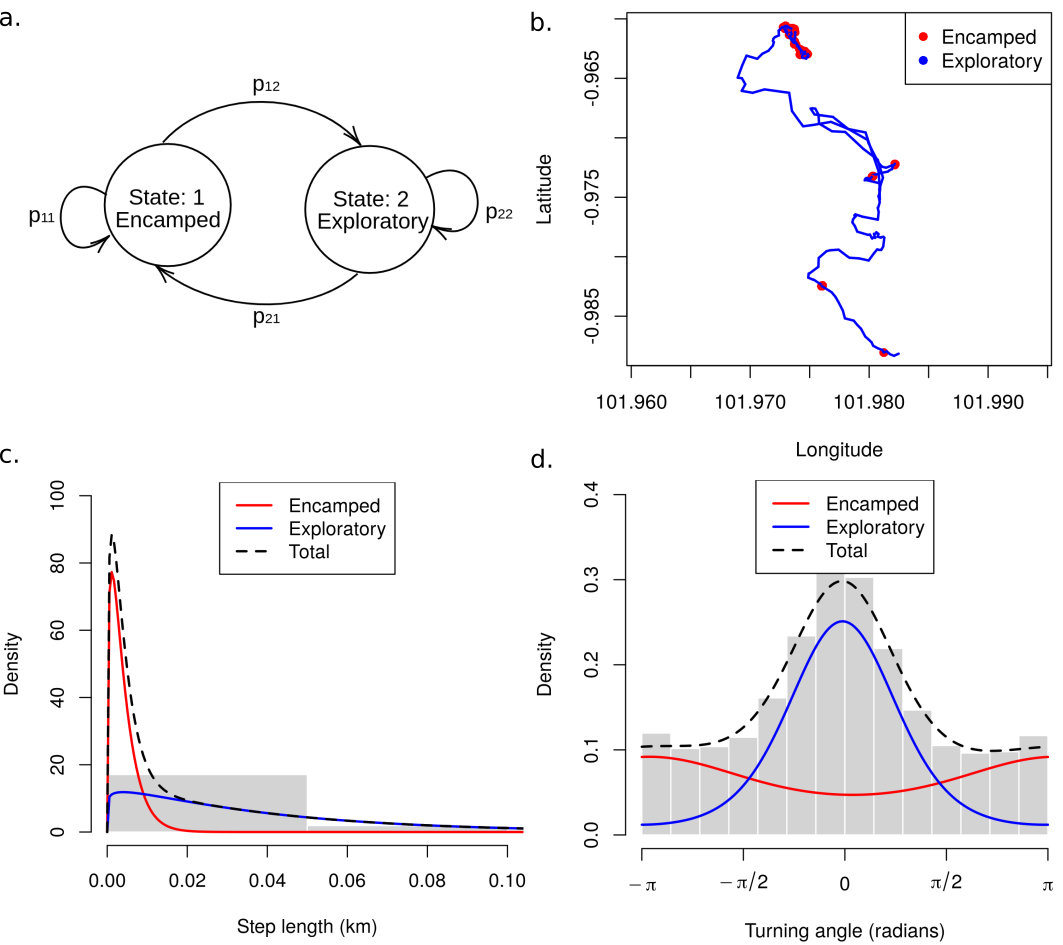}
    \caption{(a) Two-state Hidden Markov Model used to fit relocation data. $p_{ij}$ represents the state-transition probabilities where $i,j$ are the current and next states. (b) A trajectory sequence from the relocation data showing the fitted states: \textit{exploratory} states in blue and \textit{encamped} states in red. (c) Fitted Gamma distributions to the step lengths. (d) Fitted von Mises distributions to the turning angles. The histogram depicting the corresponding statistics of the relocation data is shown in gray in panels (c) and (d). }
    \label{fig:two_state_MM}
\end{figure*}

Relocation data from Asian elephants in Jambi, Indonesia (Sect.~\ref{sec:movement_data}) were used to calibrate the model parameters to match the movement statistics of the elephant agent. Ideally, we would have liked to use data from adult male elephants in the Periyar-Agasthyamali complex, but such data are not available. Therefore, we used Indonesian data as a substitute. Despite differences in location and gender, the Indonesian data suffice to identify the parameters of the stochastic movement sub-model for our ABM. The elephant agent movement model operates in discrete time, with each step defined by a \textit{step-length} and a \textit{turning-angle}. As mentioned in Sect.~\ref{section:sub-models}, two movement patterns were considered: \textit{exploratory} and \textit{encamped} walks (Fig. \ref{fig:two_state_MM} (a)). The \textit{step-length} and \textit{turning-angle} distributions were identified by fitting a two-state HMM with \textit{exploratory} and \textit{encamped} states. The best-fit distributions (Table \ref{tab:movement-parameters}) were determined using the Akaike Information Criteria (AIC) from candidate distributions (Gamma, exponential and Weibull for \textit{step-length}, and von Mises and wrapped Cauchy for \textit{turning-angle}). The gamma distribution was the best fit for \textit{step-length} (Fig. \ref{fig:two_state_MM} (c)), and the von Mises distribution was the best fit for \textit{turning-angle} (Fig. \ref{fig:two_state_MM} (d)). This analysis was performed using \textit{moveHMM}, an R package to model animal movements using HMMs.

\begin{table*}[h!]
\caption{The parameters of the fitted two-state HMM} \label{tab:movement-parameters}
\begin{tabular}{p{0.25\textwidth}p{0.15\textwidth}p{0.06\textwidth}p{0.05\textwidth}}
\toprule
movement statistic & parameter & value & unit\\ 
\midrule
State transition probability & $p_{11}$ & 0.8775 & \\
 & $p_{12}$ & 0.1225 & \\
 & $p_{22}$ & 0.9096 & \\
 & $p_{21}$ & 0.0904 & \\
 \midrule
\textit{step-length}: \textit{encamped} & mean & 0.0040 & km\\
& standard deviation & 0.0034 & km \\
\midrule
\textit{step-length}: \textit{exploratory} & mean & 0.0398 & km\\
& standard deviation & 0.0378 & km \\
\midrule
\textit{turning-angle}: \textit{encamped} & mean & -3.0232 & radians\\
& concentration &  0.3336 &  \\
\midrule
\textit{turning-angle}: \textit{exploratory} & mean & -0.0366 & radians \\
& concentration & 1.5202 &  \\
\bottomrule
\end{tabular}
\end{table*}
\subsubsection{Genetic algorithm to identify parameters in the ABM}
\label{sec:calibrating_forest_model}
Since the detailed movement data for elephants in the simulation domain, such as monthly area usage and daily travel distances were unavailable, we used the relocation data from Jambi, Indonesia to calibrate four parameters/state-variables of the ABM: \textit{prob-food-forest}, \textit{max-food-value-forest}, \textit{percent-memory-elephant} and \textit{radius-food-search}. The parameters \textit{prob-food-forest} and \textit{max-food-value-forest} capture the spatial distribution and quantity of food resources within the forest. The state variables \textit{percent-memory-elephant} and \textit{radius-food-search} is used by the elephant agent in making foraging decisions. Changing these parameters leads to different space-use patterns. Thus, an inverse problem was set up to identify the calibrated parameters that lead to the same space use patterns as in the relocation data.

Three specific spatial patterns were used as the objective functions for the calibration task: monthly space use measured by the Minimum Convex Polygon (MCP), the total daily distance traveled (diel displacement) and the daily net distance (distance between the first and last locations each day). A multi-objective genetic algorithm, Non-dominated Sorting Genetic Algorithm (NSGA) - II, was used for the calibration. The cost function was set in such a way that it penalized simulations that fell outside the confidence interval of the movement data. To simplify the calibration process, \textit{thermoregulation-threshold} was set above the ambient temperature so that the elephant agent only participates in foraging. Thus, the forest-related ABM parameters could be calibrated for food availability. As the landscape in Jambi was flat, a relatively flat elevation matrix was used during this calibration step. The ABM was further modified and adapted to incorporate movement in hilly terrains as described in Sect.~\ref{sec:calibrating_movement_wrt_slope}.

The genetic algorithm returned a pareto-optimal solution, where no alternative solution can improve one objective without compromising another. The process of selecting the final set of calibrated parameters from the pareto-optimal solution involved a critical assessment of the practical implications and feasibility of each parameter choice, with the aim of ensuring that the selected parameters align with the overall objectives and constraints of the optimization problem. The final values chosen for the parameter \textit{prob-food-forest} is 0.1 and \textit{max-food-value-forest} is 40~$kg$. The calibrated values of the elephant agent's state variables are given in Table \ref{tab:model_calibration}.

\subsubsection{Cost of movement to traverse the slopes}
\label{sec:calibrating_movement_wrt_slope}

A feasible movement direction submodel (Sect.~\ref{section:sub-models}) was required since the relocation data was obtained from a relatively flat terrain and our study area was in a hilly terrain. The primary difference in the decision making of elephant agents in these two simulation environments lies in the number of candidates for the next feasible movement direction. In the case of a flat terrain, there are a greater number of options for the possible direction of movement compared to a hilly terrain. Thus, a restrictive movement cost was introduced to identify the next feasible movement direction. 

The parameter \textit{terrain-radius} in the movement cost function was set as the calibrated \textit{radius-food-search} (Sect.~\ref{sec:calibrating_forest_model}). This was done to ensure that all the movement targets remain within the same spatial extent of the agent's current location in the elephant agent's every decision making. The parameter \textit{tolerance} was adjusted by exploratory analysis to ensure that the elephant agents traverse slopes greater than $30^\circ$ in not more than 1\% of the simulated trajectories. The inclusion of this concession was intended to avoid overly restrictive limitations and to acknowledge inherent unpredictability in the decision-making process to align with the realities of the real world.
 
Interestingly, the addition of slope-based movement cost resulted in similar metrics of MCP space usage, diel displacement, and net displacement as the movement data for a time frame of 30 days. This further justifies our use of Jambi data to tune the ABM in Seethatode.

\subsubsection{Literature-based initialization and parameterization}
\label{sec:calibrating_using_literature}

The following parameters and state variables were calibrated or initialized using an extensive literature survey. 

\paragraph{Body weight and age of elephant agents:}

The von Bertalanffy functions have been extensively used to model the growth of vertebrates in size (height, length, weight; \citep{sukumar1988a}). The general form of the equation is as follows. 

\begin{equation}
S_t = S_x (1-exp(-K(t - t_0))^M\,,
\label{eqn:body-weight-01}
\end{equation}

where, $S_t$ is the size at age $t$, $S_x$ is the asymptotic size, $K$ is the catabolism coefficient (constant), $t$ is the age of the animal in years, $t_0$ is the theoretical age at which the animal would have zero size, and $K$ is the power of the function.

This equation is assumed to be cubic for the growth of body weight ($M = 3$). We use the parameters of Eq.~\ref{eqn:body-weight-01} as determined from captive Asian elephants \citep{sukumar1988a}: $M=3, K=0.149, t_0=-3.16, S_x=4000$. In all experiments, we used a 40-year-old bull elephant. Thus, its weight according to Eq.~\ref{eqn:body-weight-01} is 4000 kg.


\paragraph{Daily dry matter intake of the elephant agents:}

The daily dry matter intake of wild Asian elephants was estimated to be 1.5\% to 1.9\% of body weight \citep{Ullrey1997}. In all experiments, we used the mean value of 1.7\% of body weight as the daily dietary requirement of the elephant agents (68 $kg/day$).  

\paragraph{Spatio-Temporal thermoregulation probability:}

In this study, spatio-temporal temperature data were collected from the WorldClim data website [CRU-TS 4.06 \citep{Harris2020}]. In particular, maximum and minimum monthly temperature data were collected at a spatial resolution of 2.5 minutes for the corresponding simulation year. Since we are interested in the daily movements of elephants, the minimum and maximum climatological temperature was used with the empirical daily temperature curve to obtain hourly temperature data for the simulation time period (This was done using \textit{chillR} in R). The \textit{prediction\_coefficient} needed in the empirical temperature curve was used from the nearest weather station closest to the study area (Thiruvananthapuram). 

\begin{figure}[!h]
    \centering
    \includegraphics[width=0.65\textwidth]{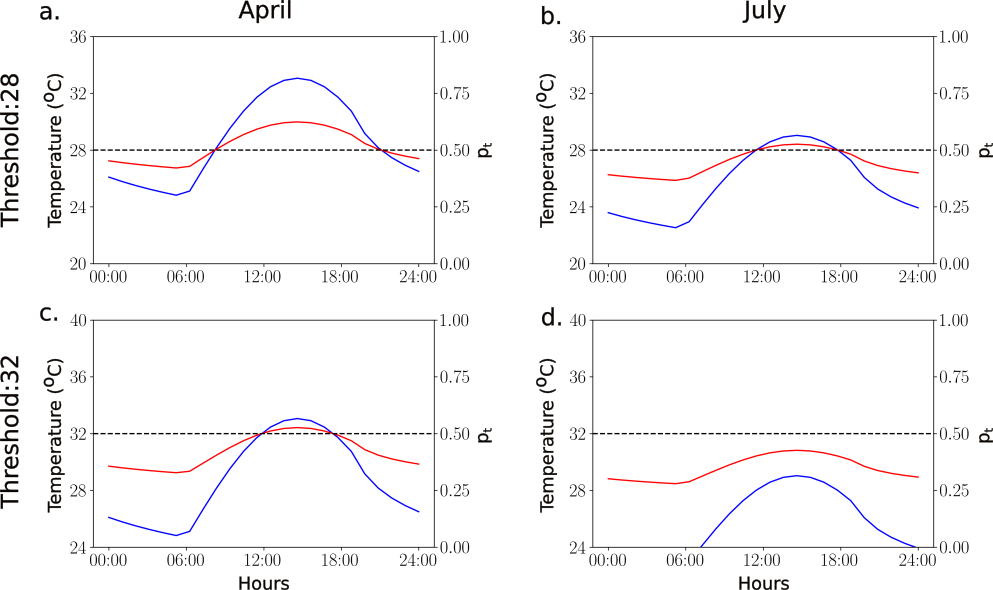}
    \caption{The figure illustrates the daily temperature variation (blue line) in the landscape cell at the center of the simulation area and the corresponding probability of thermoregulation ($p_{t}$) (red line) for a dry month (April) and a wet month (July) of the year. The horizontal line represents \textit{$T_{threshold}$}. The left y-axis shows temperature ($^\circ C$) and the right y-axis shows the probability of thermoregulation.}
    \label{fig:daily_p_t_pattern}
\end{figure}

Given a threshold for thermoregulation, the probability of thermoregulation was calculated from the hourly temperature using the following formula \citep{Diaz2021}: 

\begin{equation}
    p_{t} = \frac{1} {1+ \exp(state \times [T_{current} - T_{threshold}]}
\label{thermoregulation_probability}
\end{equation}

where, $p_{t}$ is the \textit{thermoregulation-probability}, \textit{ state} may be -0.2 or -0.1, highlighting the difference in sensitivity to environmental temperatures between family groups with and without calves, \textit{$T_{current}$} is the temperature of the agent's landscape cell, and \textit{$T_{threshold}$} is the temperature threshold for thermoregulation. The state was set as $-0.1$ for \textit{solitary bulls} and $-0.2$ for \textit{matriarchal herds}. If the \textit{thermoregulation-probability} ($p_{t}$) was greater than 0.5 in its current landscape cell, the elephant agent would engage in thermoregulation. In the experiments, two \textit{thermoregulation-threshold}s of $28^\circ$ and $32^\circ$ were chosen so that the elephant agent thermoregulates in all seasons at threshold $28^\circ$ and only during the dry/hot season at the threshold $32^\circ$ (see Fig. \ref{fig:daily_p_t_pattern}).

\subsubsection{Other assumptions used in the calibration of the model}
\label{calibrating_others}

Based on the data collected from the questionnaire (Sect.~\ref{sec:field_survey_and_data_collection}), conflict reports were recorded at a distance of 1500~$m$ from the fringe. Therefore, the \textit{knowledge-from-fringe} variable was set to 1500~$m$.

It is assumed that the elephant agent will die in 10 days if it does not meet its food and thermoregulation needs. This would mean that each day \textit{fitness} decreases by a factor of 0.1. The simulation runs at a temporal resolution of 5~minutes and each day comprises 288 time steps. Therefore, in each time step, \textit{fitness} is reduced by a factor of 0.000347. This energy expended must be restored by eating food and maintaining a stable body temperature by thermoregulation. However, thermoregulation demands fluctuate throughout the seasons. As a result, depleted \textit{fitness} must be increased daily by the values of \textit{fitness-increment-when-thermoregulates} and \textit{fitness-increment-when-eats-food}, proportionate to the number of time steps the agent spends in thermoregulation in a day. 

Let \textit{a} be the total number of thermoregulation time steps in a day; $x$ be the total food consumed by the elephant agent in a day ($kg$); $y$ be the total number of time steps that the agent actually thermoregulated in a day.

\begin{equation}
\begin{split}
\textit{fitness-increment-when-eats-food} = \frac{1}{10} \times \frac{288-a}{288} \times \frac{\min(x, \textit{daily-dry-matter-intake})}{\textit{daily-dry-matter-intake}}
\end{split}
\label{eqn:fitness-increment-when-eats-food}
\end{equation}

\begin{equation}
\begin{split}
\textit{fitness-increment-when-thermoregulates} = \frac{1}{10} \times \frac{a}{288} \times \frac{y}{a}
\end{split}
\label{eqn:fitness-increment-when-thermoregulates}
\end{equation}

Here, $\frac{1}{10}$ represents the proportion of daily \textit{fitness} loss, $\frac{288-a}{288}$ represents the proportion of the day not thermoregulating (available for eating), $\frac{max(x, daily-dry-matter-intake)}{daily-dry-matter-intake}$ promotes eating more than the daily requirement. If the elephant eats more than \textit{daily-dry-matter-intake}, it receives the full benefit. Otherwise, the benefit scales proportionally to the amount eaten. $\frac{a}{288}$ represents the proportion of the day spent thermoregulating, and $\frac{y}{a}$ represents the efficiency of thermoregulation. If $y$ is equal to $a$ (perfect efficiency), full benefit is received. Otherwise, the benefit scales proportionally to the actual time spent thermoregulating. If the agent's fitness level is lower than its maximum fitness and it consumes more than its dietary needs, it gains a fitness advantage because of the excess food consumed. This benefit is scaled according to the surplus food consumed during the day.

The \textit{movement-fitness-deprecation} is decreased from \textit{fitness} at each time step, while \textit{fitness-increment-when-eats-food} and \textit{fitness-increment-when-thermoregulates} is updated at the end of every day. 

The parameter \textit{fitness-threshold} was set such that the elephant agent focuses on foraging if the energy falls below 40\% of its maximum value. 

The value of the parameter \textit{radius-water-search} was set to be equal to the value of \textit{radius-food-search} for replicating spatial patterns of the collected movement data.
\begin{table*}
\caption{Summary of the calibrated state variables of the elephant agent.}
\label{tab:model_calibration}
\resizebox{0.985\textwidth}{!}{
    \begin{tabular}{p{0.2\textwidth}p{0.075\textwidth}p{0.30\textwidth}p{0.075\textwidth}p{0.320\textwidth}}
    \toprule
    \textbf{Parameter} & \textbf{Value} & \textbf{Assumptions} & \textbf{Paper Sect.} & \textbf{Supporting references} \\
    \midrule
    \textit{age} & 40~years & The experiments involved adult solitary male elephant & 2.4.5 & \\
    \midrule
    \textit{body-weight} & 4000~kg & & 2.4.4 & von Bertalanffy function for the body weight of male elephants \citep{sukumar1988a} \\
    \midrule
    \textit{daily-dry-matter-intake} & 68~kg & 1.7\% of body weight & 2.4.4 & 1.5\% to 1.9\% of body weight \citep{Ullrey1997} \\
    \midrule
    \textit{knowledge-from-fringe} & 1500~m & Parameterized using collected field data & 2.4.5 & Records of damage incidents within a buffer zone close to the fringe \citep{Kalyanasundaram2013, Max2010}\\
    \midrule
    \textit{percent-memory-elephant} & 0.375 & Replicating spatial patterns of collected movement data & 2.4.2 & \\
    \midrule
    $p_{t}$ & & $1/[1+ \exp(state \times (t_{current} - t_{threshold})]$ & 2.4.4 & probability of thermoregulation given a threshold temperature \citep{Diaz2021} \\
    \midrule 
    \textit{thermoregulation-threshold} $(T_{threshold})$ & $28^{\circ}C$,$32^{\circ}C$ & Thresholds chosen so that the elephant agent thermoregulates in all seasons and only during dry season & 2.4.5 &  \\
    \midrule
    \textit{radius-food-search} & 750~m & Replicating spatial patterns of collected movement data & 2.4.2 & \\
    \midrule
    \textit{radius-water-search} & 750~m & Replicating spatial patterns of collected movement data & 2.4.5 & \\
    \midrule
    \textit{radius-forest-search} & 1500~m & Knowledge of forest from fringe if close to the fringe & 2.4.5 & Due to conflict, elephants are forced to venture deeper into human settlements, thus facing challenges in their ability to retreat back to forests \citep{Kalyanasundaram2018} \\
    \midrule
    \textit{movement-fitness-deprecation} & 0.000347 & If the elephant agent cannot meet its thermoregulation or food requirements, it will die in 10 days & 2.4.5 & \\
    \midrule
    \textit{fitness-increment-when-eats-food} & & $[(288-a)\times[\max(x, \textit{daily-dry-matter-intake})]] /(288 \times \textit{daily-dry-matter-intake})$ & 2.4.5 & \\
    \midrule
    \textit{fitness-increment-when-thermoregulates} & & $$\frac{1}{10} \times \frac{a}{288} \times \frac{y}{a}$$ & 2.4.5 & \\
    \midrule
    \textit{fitness-threshold} & 0.4 & Focus on \textit{foraging} if the energy falls below 40\% of its maximum value & 2.4.5 & \\
    \midrule
    \textit{terrain-radius} & 750~m & same as the radius of search of the food and water source & 2.4.3 & \\
    \midrule
    \textit{tolerance} & 100 & & 2.4.3 &  \\
    \bottomrule
    \end{tabular}
}
\end{table*}

\section{Model validation}\label{Model_validation}
\begin{figure}[h!]
    \centering
    \includegraphics[width=0.485\textwidth]{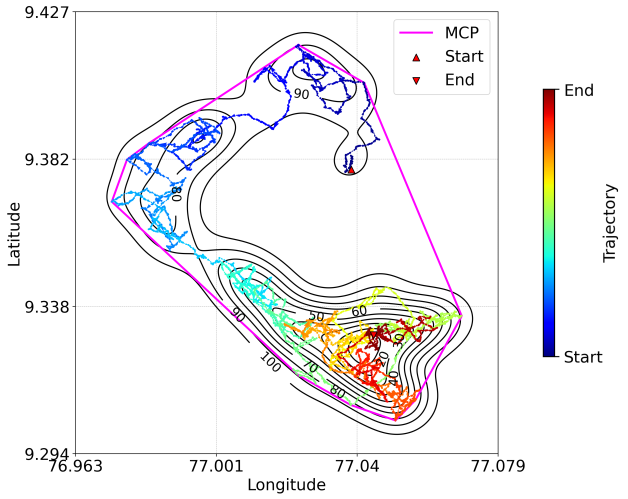}
    \caption{A simulated 30-day elephant trajectory with MCP and KDE shown.}
    \label{fig:area_usage}
\end{figure}

Model validation is conducted through both qualitative and quantitative comparisons with patterns documented in the literature. Table \ref{tab:model_validation} lists the patterns from the literature that are accurately replicated by the ABM. Fig.~\ref{fig:area_usage} illustrates the Minimum Convex Polygon (MCP) and the Kernel Density Estimate (KDE) of a simulated path. The MCP area is calculated to be 77.65~$km^2$. The KDE contours fitted to various levels of space utilization are also depicted. The region representing 100\% space use encompasses 57.82~$km^2$, while 90\% space use corresponds to 36.68~$km^2$. These values align well with those reported in the literature. Additionally, there is a strong agreement with the conflict data gathered from the field.

\begin{table*}[h!]
\caption{Validation of ABM}\label{tab:model_validation}
\begin{tabular}{p{0.45\textwidth}p{0.45\textwidth}}
\toprule
Patterns documented in the literature & Emergent patterns from the ABM \\ 
\midrule
Activity budgets & \\
{} & {} \\
An inverse relationship was observed between increasing temperatures and feeding activity, while a direct relationship was found between resting and increasing temperatures \citep{Baskaran2010}. & With increasing temperatures, the elephant agent rests near a water source or within the forest shade. \\
\midrule
Space usage & \\
{} & {} \\
During the dry season, elephants congregated in river habitats, while they dispersed during the wet season \citep{Sukumar1989}. & The dry season leads to congregations in river valleys, while the dispersion patterns during the wet season are influenced by the availability of food in forested areas and plantations (see Fig. \ref{fig:dry_wet_months_density_plots}). \\
{} & {} \\
The spatial usage measured by the MCP was 35~$km^2$ in the dry season and 50~$km^2$ in the wet season \citep{AnandaKumar2010}. & The area utilization, determined by the MCP and KDE, exhibits comparable values (see Fig. \ref{fig:area_usage}). \\
{} & {} \\
The elephants exhibited a shuttling movement pattern when visiting water sources \citep{thaker2019}. & The model replicated the shuttling behaviors of the elephants with respect to water source visit (see Fig. \ref{fig:day_night_trajectories}). \\
{} & {} \\
Elephants tended to be close to water sources during the warmer parts of the day and spent more time at water sources during the dry season than during the wet season \citep{thaker2019}. & The model replicated similar patterns of space use diurnally and during the wet and dry months (see Fig. \ref{fig:day_night_trajectories}). \\
\midrule
Conflict patterns & \\
{} & {} \\
Approximately 60\% of the damage occurred close to the edges of the forest, within a distance of no more than 50~$m$ from the boundary \citep{Rohini2015}. & The conflict incidents were located mainly in close proximity to the forest-plantation boundary (see Fig. \ref{fig:dry_wet_months_density_plots}). \\
\bottomrule
\end{tabular}
\end{table*}

\section{Experimental setup}
\label{experimental_setup}

We conducted several experiments to explore the spatio-temporal space use patterns of elephant agents along three major axes: food availability conditions, elephant aggression levels, and thermoregulation thresholds. The simulation experiments were conducted for 2010 corresponding to the questionnaire survey field data. All experiments start on the first day of a month and end on the last day of that month. The calibrated state variables and model parameters are used for the experiments. In all experiments, one bull elephant agent with the same initial location (Sect.~\ref{section:initialization}) is used. For each month, 192 simulations are performed to obtain trajectories, food consumption, crop raiding incidents, building damage incidents, movement and space use statistics. 

\paragraph{Effect of food availability.} The first inquiry axis to study the movement and spatial utilization patterns of elephants is different food availability conditions. We have estimated food availability and needs based on the dry matter requirement for elephants in every scenario. In the current prototype ABM, the growth and decay of crops are not included during the simulation period. The \textit{food-matrix} serves as the primary input for the elephant agent in the ABM to make movement decisions for food consumption. This matrix represents two key factors: the spatial distribution and the quantity of food resources. Only forest cells and plantation cells contain food. The presence of food in each forest cell (or plantation cell) is considered an Independent and Identically Distributed (IID) binomial random variable with a probability \textit{forest-food-percent} (or \textit{cropland-food-percent}). If a cell contains food, the amount of food available in that cell is a uniform random variable between 0 and \textit{forest-max-food-value} (or \textit{cropland-max-food-value}). 

To determine the distribution of food in plantation cells (\textit{cropland-food-percent}), we used field questionnaire data (Sect.~\ref{sec:field_survey_and_data_collection}) on agricultural plots in the study area. This sample covers 1.65 sq km and represents the land use of plantation cells in the simulation domain. Data reveal that rubber dominates the survey area, with home gardens scattered throughout, providing food for elephants. Plantation cells can contain both rubber and home gardens, classified by percentage of rubber canopy (Appendix \ref{sec:classification_agricultural_plots}). The three main categories are: 75-100\% rubber canopy with home gardens (36.9\%), 50-75\% rubber canopy (29.6\%), and 25-50\% rubber canopy (15.7\%). Other plot types are insignificant. We assign each plantation cell to one of the seven survey categories according to their area distribution (Fig.~\ref{fig:agricultural-plot-distribution}). Not all cells have food for elephants and only home gardens are raided. We designate plantation cells as home gardens if their category includes home gardens, ensuring that about 35\% of the plantation area are home gardens, according to the survey data. For example, if 100 cells are assigned a 50-75\% rubber canopy, 37 (37.5\%) are designated as home gardens with potential food for elephants. 



Estimating the food supply for elephants in plantations is complicated by various sources of uncertainties. The home gardens could have various species of trees, not all edible by elephants. In addition, not all of these species produce fruits at the same time. Although we lack specific details, we know that certain species have seasonal fruiting periods. We are also uncertain about the specific types of trees and the parts of these trees that elephants can and do consume within the study area. Another challenge is that the survey only provides information about the presence or absence of a specific tree species in a certain category of the agricultural plot, not their count or growth stage. We address this uncertainty as follows. Based on reported crop raid incidents, we assume that elephants prefer the following species: "Plantain" (\textit{Musa paradisiaca}), "Tapioca" (\textit{Manihot esculenta}), "Cashew" (\textit{Anacardium occidentale}), "Tamarind" (\textit{Tamarindus indica}), "Jackfruit" (\textit{Artocarpus integrifolia}), "Mango" (\textit{Mangifera indica}), "Rambuthan" (\textit{Nephelium lappaceum}), and "Guava" (\textit{Psidium guajava}). Of these, "Plantain" is a perennial crop, while the others are seasonal. We assume that each landscape cell can contain multiple types of tree species and multiple counts of the same tree species fruiting in different months. A Bernoulli random variable was used to model the presence of food in a plantation cell in any given simulation month with a probability \textit{cropland-food-percent}. We know that a mature plantain typically bears fruit over a 3- to 4-month period. Averaging this to 3.5 months, it can be inferred that there is a monthly probability of 0.3 for the presence of fruits in a landscape cell with plantain. We also assumed that 1/12 of cells with a home garden and no plantains bear fruit in any given month. Combined, we arrive at \textit{cropland-food-percent}=0.3.

Research indicates that elephant crop raiding can be viewed as an extension of their foraging behavior, targeting cultivated cereal and millet crops because of their higher nutritional content compared to wild grasses \cite{Sukumar1990}. Crops like finger millet have higher levels of protein, calcium, and sodium than wild plants, which may enhance survival, growth, and physical condition, particularly aiding in male-male competition \cite{Sukumar1988b}. Although the crops mentioned in the literature are not prevalent in the simulation domain, we use insights from the literature to set model parameters. In our experiments, we distinguish between the quality of forage in forested areas and croplands without explicitly modeling caloric content, assuming that elephants can more easily meet their nutritional needs by foraging in croplands rather than forests, suggesting a greater dietary benefit from croplands. A study on crop-raiding elephants found that adult bulls typically consume about 43.9 kg of crops daily and can eat between 70-75 kg of finger millet in a single night \cite{Sukumar1990}. The parameter \textit{cropland-max-food-value} was chosen to increase the likelihood that elephant agents meet their daily dietary needs during a crop-raiding event. Such events usually occur at night when human activity is low. The event begins when an elephant agent enters a plantation cell in search of food and ends when the agent returns to the forest in the morning. Given that the food quality in plantation cells is superior to that in forests, we assumed that agents could meet their daily nutritional needs in one crop-raiding episode. To achieve this, we set \textit{cropland-max-food-value}=100~$kg$. In the current prototype ABM, these modeling choices are akin to setting a caloric requirement and establishing a conversion from food weight to calories based on the nutritional quality of the crops.

Considering that croplands and human settlements offer elephants a higher quality and quantity of food sources compared to forests, the parameters \textit{forest-max-food-value} and \textit{prob-food-forest} were set lower than their counterparts \textit{cropland-max-food-value} and \textit{prob-food-cropland}. Food availability in the simulation area is uncertain, so we employed a parameter sweep method to quantify this uncertainty and examine the response of elephant agents to various food abundance scenarios. Specifically, we used five different scenarios that ranged from scarce to abundant food availability, labeled S1, S2, S3, S4 and S5, with \textit{forest-max-food-value} values of 5, 10, 15, 20, and 25~$kg$, respectively. For S1 to S5, \textit{forest-food-percent} was set to the calibrated \textit{prob-food-forest} (Sect.~
\ref{sec:calibrating_forest_model}). Table
\ref{tab:food_matrix} details the average food availability within the forest for different \textit{forest-max-food-value} parameterizations.



\begin{table*}[!h]
     \caption{Parameter values related to food consumption. S1-S5 are scenarios used in the experiments.}
     \label{tab:food_matrix}
     \begin{tabular}{p{0.25\textwidth}p{0.4\textwidth}p{0.25\textwidth}}
     \toprule
     \textbf{Parameter} &  \textbf{Values} & \textbf{Unit} \\
     \midrule
      \textit{forest-food-percent}  & 0.1 & \\
      \textit{cropland-food-percent}  & 0.3 & \\
      \textit{cropland-max-food-value}  & 100 & kg \\
      \textit{forest-max-food-value}  & S1:5, S2:10, S3:15, S4:20, S5:25 & kg \\
      food within forest & S1:0.10, S2:0.25, S3:0.35, S4:0.45, S5:0.60 & $ton/km^2$\\
      \bottomrule
     \end{tabular}
\end{table*}

\paragraph{Effect of aggression.}
The second inquiry axis for understanding space utilization and movement behaviors is the levels of aggresion. We tested four different aggression values (0.2, 0.4, 0.6, 0.8) in our experiments. These levels of aggression indicate the probability of choosing cells near plantation cells during foraging decisions. 

\paragraph{Effect of thermoregulation threshold.}
The third axis of inquiry is the thermoregulation threshold. When the daily temperature of the agent's landscape cell exceeds the thermoregulation temperature threshold, the agent must engage in thermoregulation actions. Due to the lack of precise temperature thresholds for thermoregulation for various groups of elephants in the simulation domain, we examined two different thresholds ($28^\circ$ and $32^\circ$ C). These thresholds were selected such that the elephant agent thermoregulates throughout all seasons at $28^\circ$ C and only during the dry season at $32^\circ$ C.

\paragraph{Number of ABM simulations.}
Overall, 5 (food scenarios) $\times$ 4 (aggression levels) $\times$ 2 (thremoregulation thresholds) = 40 scenarios are simulated for each of the 12 months. In each month, 192 trajectories are obtained. Therefore, a total of 40$\times$12$\times$192 = 92,160 ABM simulations have been completed. 192 trajectories are chosen as we use four CPU nodes each with 48 cores, giving a total of 192 cores being used for the simulation. All trajectories start with identical initial conditions, and the stochasticity in the submodels leads to a multitude of possible paths that are analyzed to obtain the movement patterns. 

\section{ABM Simulation Results}\label{Results}
The prototype ABM simulations offer key insights into the seasonal impacts on elephants' daily activity budgets, their emergent space use patterns, and crop raiding patterns. The key findings for each aspect are detailed below. 
\subsection{Seasonal effect on daily activity budget of elephant agents}

\begin{figure}[h!]
    \centering
    \includegraphics[width=0.48\textwidth]{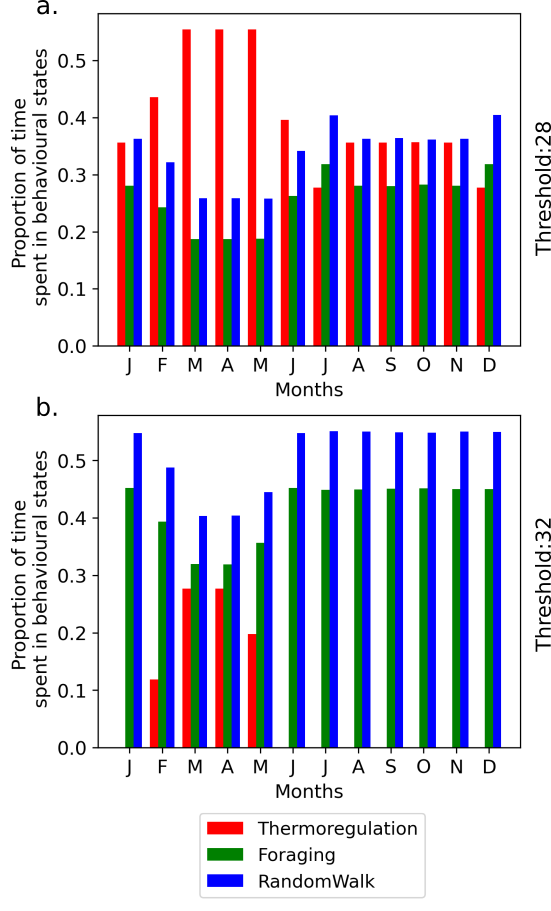}
    \caption{The agent's daily time budget across various behavioral states for different months of the year. (a) \textit{$T_{threshold}$}: 28°C (b) \textit{$T_{threshold}$}: 32°C.}
    \label{fig:state_probabilities}
\end{figure}

We analyze the patterns in the daily average time spent by the elephant agent in each of the \textit{thermoregulation}, \textit{random-walk}, and \textit{foraging} modes across the different months (Fig.~\ref{fig:state_probabilities}). The first observation is that when the thermoregulation threshold $T_{threshold}=28^\circ$C, the agent spends time in all three modes in all months; however, when $T_{threshold}=32^\circ$C, the agent enters the \textit{thremoregulation} mode only in February, March, April, and May, which is the hot/dry season. A series of chi-square tests was performed to identify pairs of months with a statistically significant difference ($\alpha=0.05$) in the distribution of the daily average time budget in the three modes. We found that the differences in the daily time budget of the agent states were statistically significant between the dry and wet months (p < 0.05) for all aggression levels at both $T_{threshold}$s. Chi-square tests were also performed to see if aggression levels and food availability within the forest resulted in differences in the distribution of the daily time budget. The results indicated that there are no statistically significant differences (p < 0.05). 

\subsection{Emergent space use patterns of the elephant agent}

\begin{figure*}[h!]
\centering
    \includegraphics[width=0.925\textwidth]{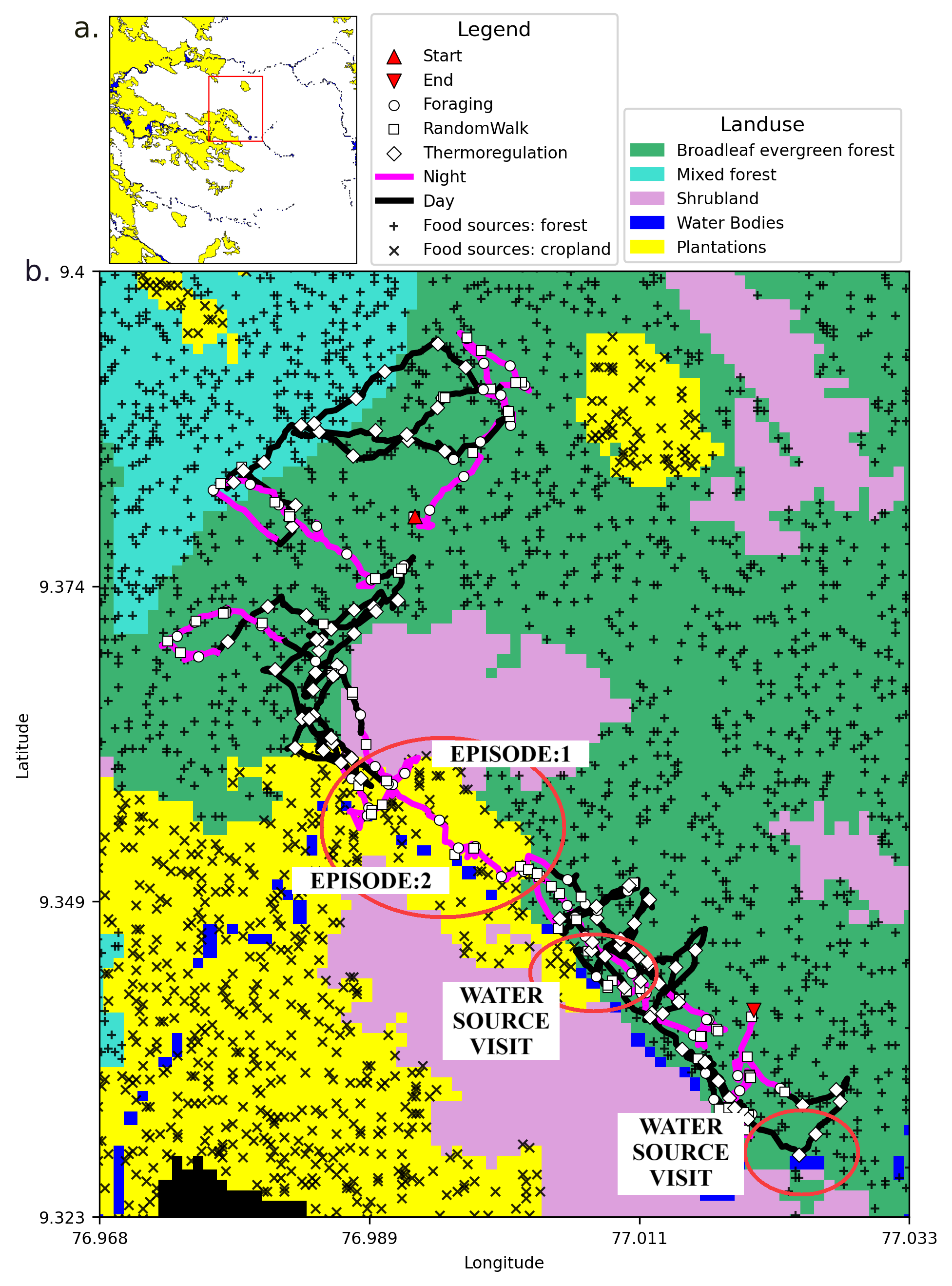}
    \caption{A simulated elephant trajectory sequence highlighting the activity during the day and night.}
    \label{fig:day_night_trajectories}
\end{figure*}

Fig.~\ref{fig:day_night_trajectories} shows a simulated trajectory sequence of an elephant agent in a dry month over eight days, demonstrating various dynamics and mechanisms that govern the elephant's movement through the landscape. Fig.~\ref{fig:day_night_trajectories}(a) shows the entire simulation area, highlighting plantations in yellow and water bodies in blue. The red box shown in Fig.~\ref{fig:day_night_trajectories}(a) highlights the zoomed area shown in Fig.~\ref{fig:day_night_trajectories}(b). Fig.~\ref{fig:day_night_trajectories}(b) presents a sequence of the elephant's movements, distinguished by daytime in black and nighttime in pink. Scatter points along the trajectory represent the elephant's behavioral mode at each hourly interval. These points represent three distinct states: \textit{thermoregulation}, \textit{random-walk}, and \textit{foraging}, indicating how the behavior of the elephant changes throughout the day. The trajectory reveals various movement dynamics, including visits to water bodies and plantations, foraging, and crop-raiding, as well as landscape navigation based on slope and elevation. Importantly, the trajectory demonstrates recurrent crop raiding events (episode:1 and episode:2 in Fig.~\ref{fig:day_night_trajectories}), both occurring at night. The trajectory sequence also highlights the shuttling movement of the simulated elephant agents with respect to the water sources when in \textit{thermoregulation} mode (events marked as `water source visit').

\begin{figure*}[h!]
    \centering
    \includegraphics[width=\textwidth]{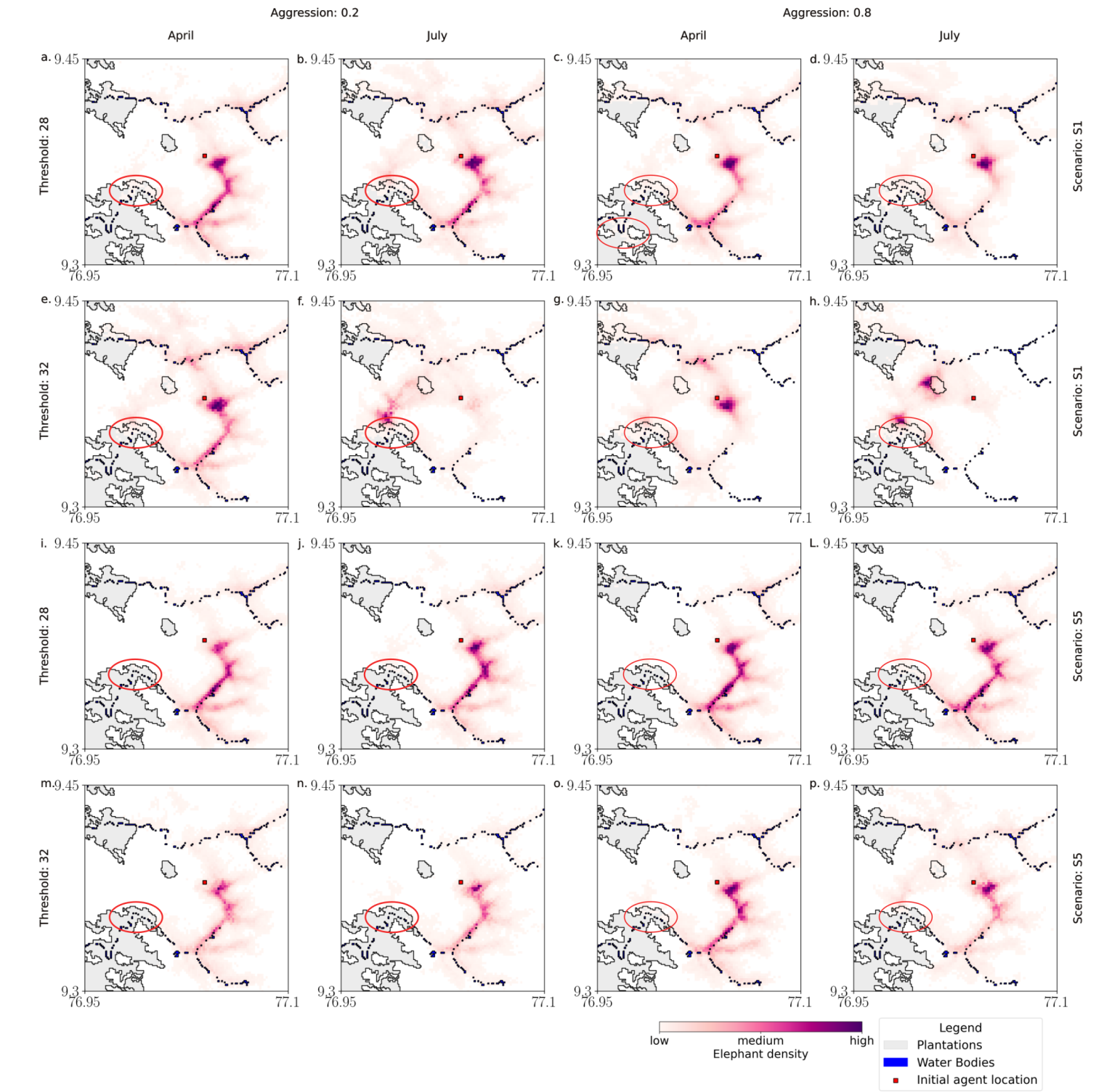}
    \caption{The summary of the spatial distribution of the agents at the two thermoregulation thresholds, two extreme aggression levels and two extreme food availability scenarios. Water sources and croplands are shown in the background. The red circles denotes the conflict clusters as identified in Sect.~\ref{sec:field_survey_and_data_collection} from the field data.}
    \label{fig:dry_wet_months_density_plots}
\end{figure*}


Fig.~\ref{fig:dry_wet_months_density_plots} shows the influence of varying levels of aggression, thermoregulation threshold, and food availability within the forest on the spatial distribution of elephants in the wet and dry months. The plots incorporate all simulated elephant trajectories to capture the inherent variability and uncertainty in the predictions of the elephant path. Dark red regions represent areas of high elephant density, whereas light red areas indicate lower densities under simulated conditions. The spatial distribution of the elephant agents is superimposed relative to the water sources and croplands within the study area. 

The regions of documented human-elephant conflicts (Sect.~\ref{sec:field_survey_and_data_collection}, Cluster I, II, and III in Fig.~\ref{fig:study_area_detailed}) are highlighted with red ovals in Fig.~\ref{fig:day_night_trajectories}(b). The simulations indicated that crop raiding incidents in cluster I occurred in all experimental setups, with variations influenced by the availability of forest food, aggression levels, and thermoregulation thresholds. Incidents in cluster II were only observed during the dry months with the highest aggression levels and a thermoregulation $T_{threshold}$ of 28$^\circ$C. This implies that elephants ventured further into croplands (cluster II) when: \textit{(i)} they needed to spend a significant portion of their daily time (over 50\%) on thermoregulation, \textit{(ii)} they exhibited high aggression, and \textit{(iii)} there was a severe shortage of food in the forest. The combined effects of demands for thermoregulation, aggression, and potential food scarcity in the forest likely drove these excursions. In addition, the presence of water along with food may have influenced their migration to these areas, potentially satisfying both needs. However, only aggressive elephants reached these areas, probably because of their higher tolerance to the risk of venturing deeper into croplands. This analysis reveals a complex interaction between food availability, seasonal changes, and aggression in influencing elephant crop raiding behavior. Finally, no incidents were observed in cluster III during any simulation. This could be due to terrain and the fact that our simulations used only a single agent starting from the same location. Future work could involve more agents and longer simulations to identify the factors leading to conflicts in cluster III.

\begin{figure}[h!]
    \centering
    \includegraphics[width=0.625\textwidth]{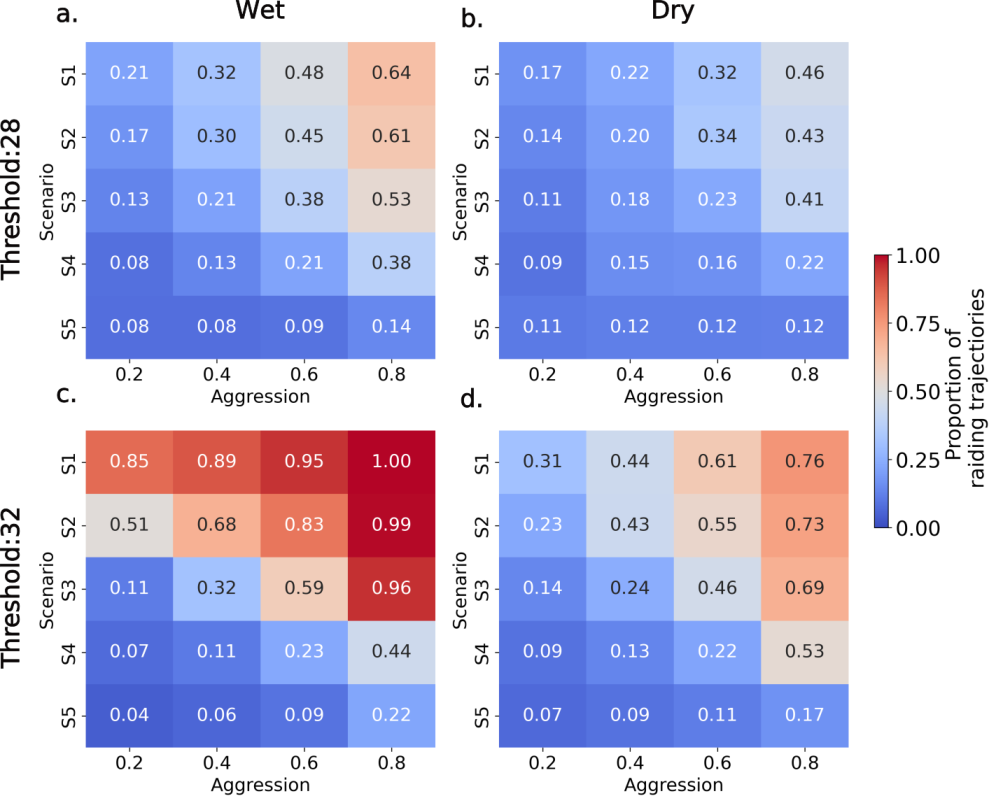}
    \caption{The variation in the proportion of the simulated elephant trajectories that exhibit crop-raiding behavior with aggression, forest food availability and thermoregulation. This proportion can be viewed as the probability of conflict.}
    \label{fig:prop_traj_raiding}
\end{figure}

\begin{figure}[h!]
    \centering
    \includegraphics[width=0.625\textwidth]{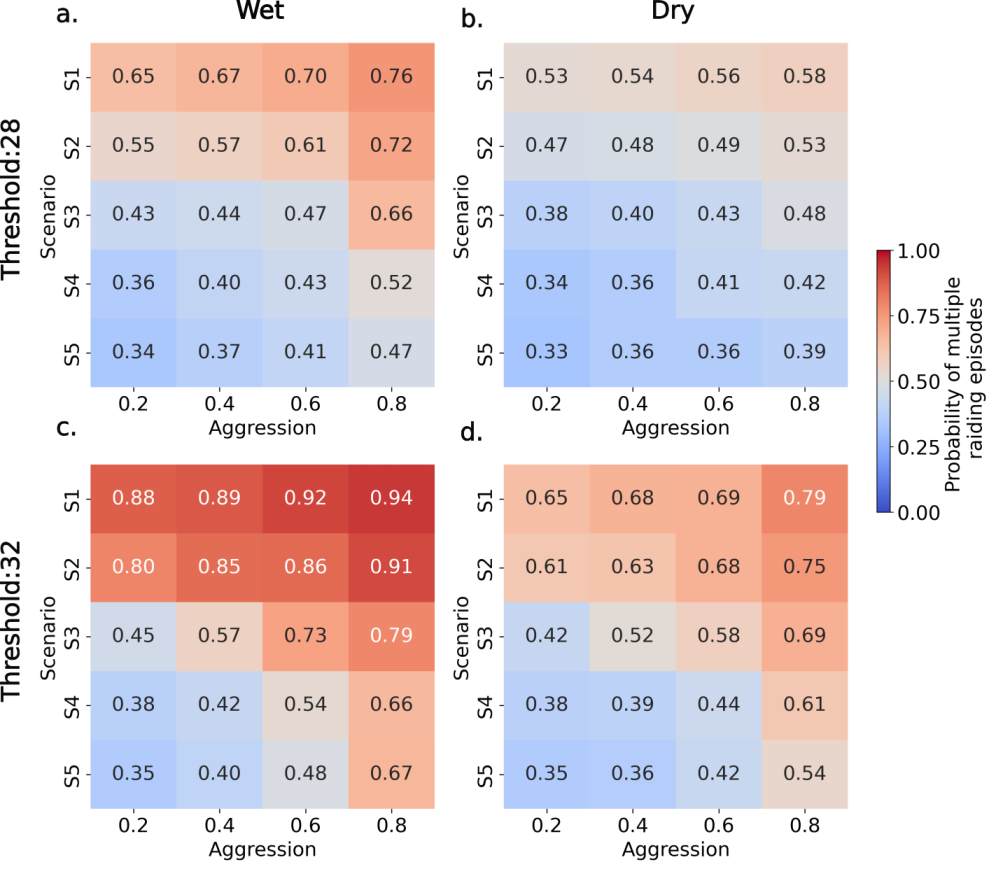}
    \caption{The likelihood of making multiple returns to the plantations by the crop-raiding elephant agents.}
    \label{fig:prop_reentry}
\end{figure}

\subsection{Human-elephant conflict: Crop raid patterns}

\begin{figure}[h!]
    \centering
    \includegraphics[width=0.625\textwidth]{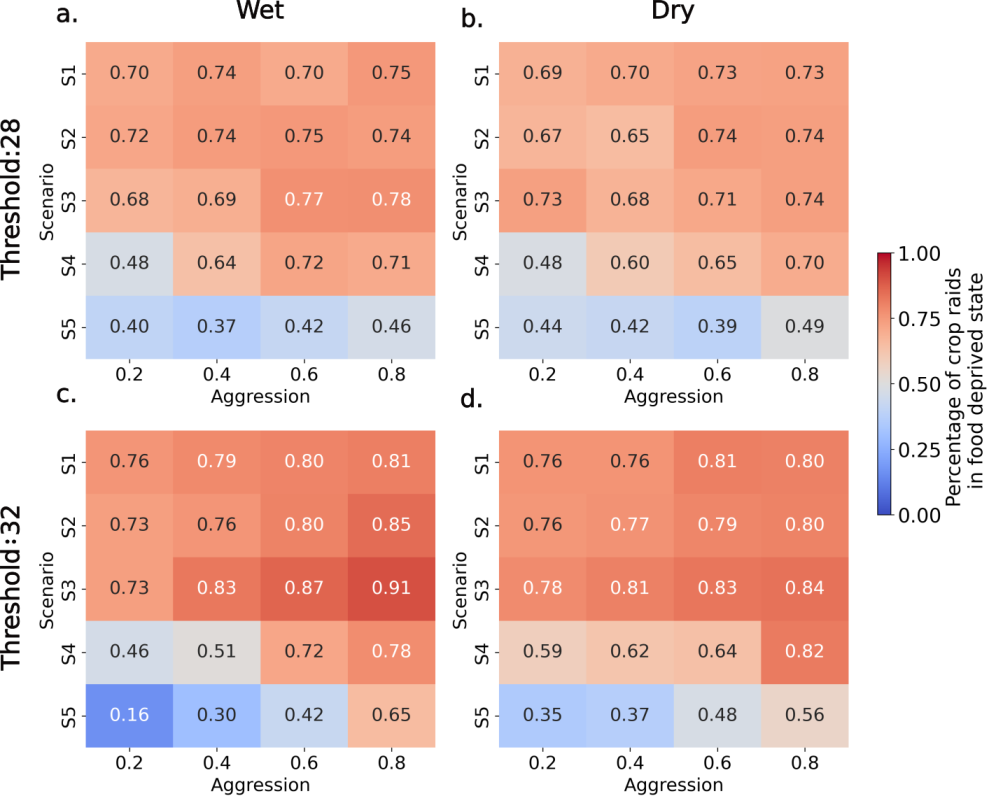}
    \caption{The proportion of crop raid episodes in which elephants are in a food-deprived state.}
    \label{fig:prop_incidents_deprecated}
\end{figure}

\begin{figure*}[h!]
    \centering
    \includegraphics[width=0.925\textwidth]{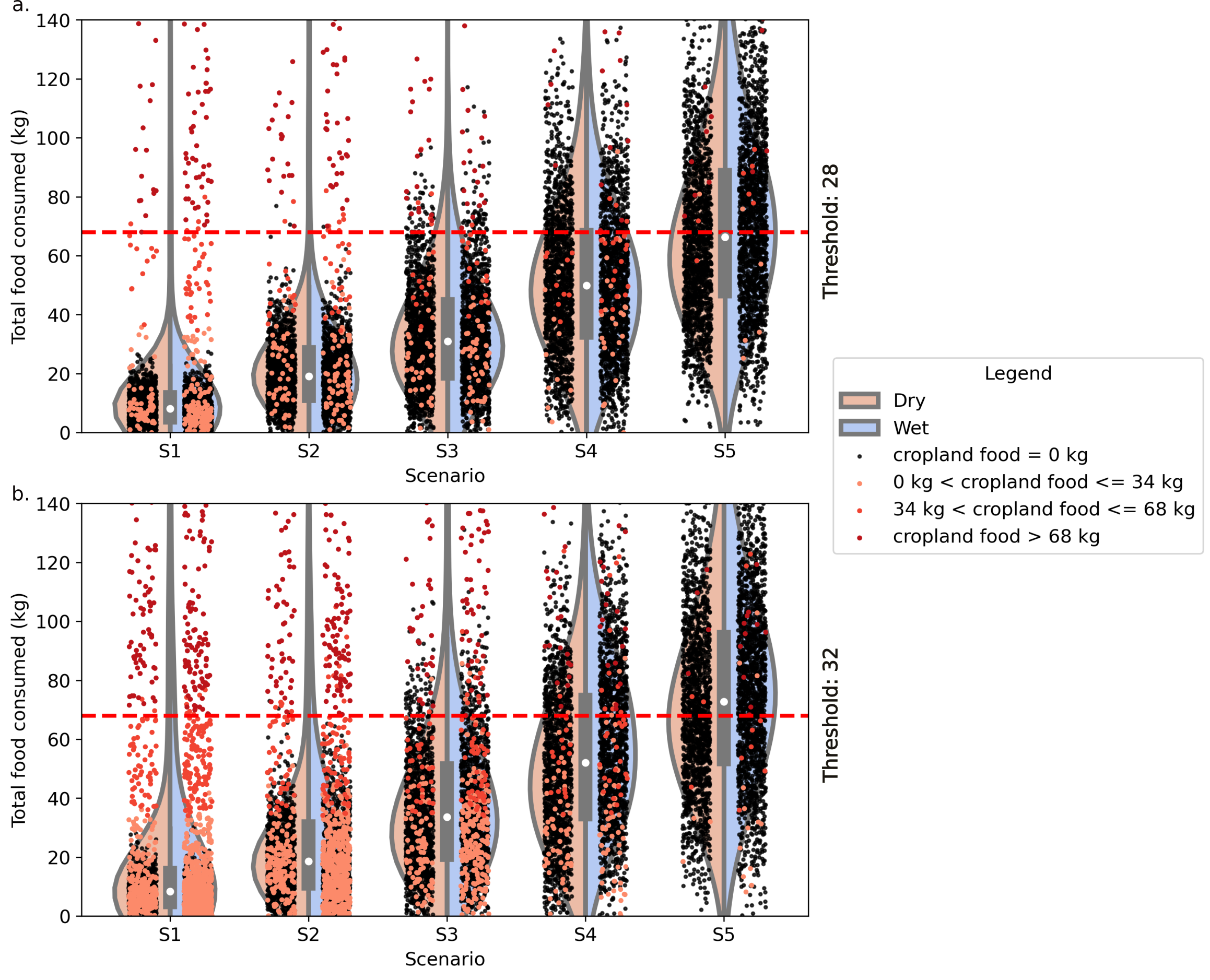}
    \caption{The daily food consumption of the elephant agents under various scenarios. The distribution is obtained considering samples of an equal number of days (5000 days) from the simulated trajectories for the wet and dry months.}
    \label{fig:food_distributions}
\end{figure*}

\begin{figure*}[h!]
    \centering
    \includegraphics[width=\textwidth]{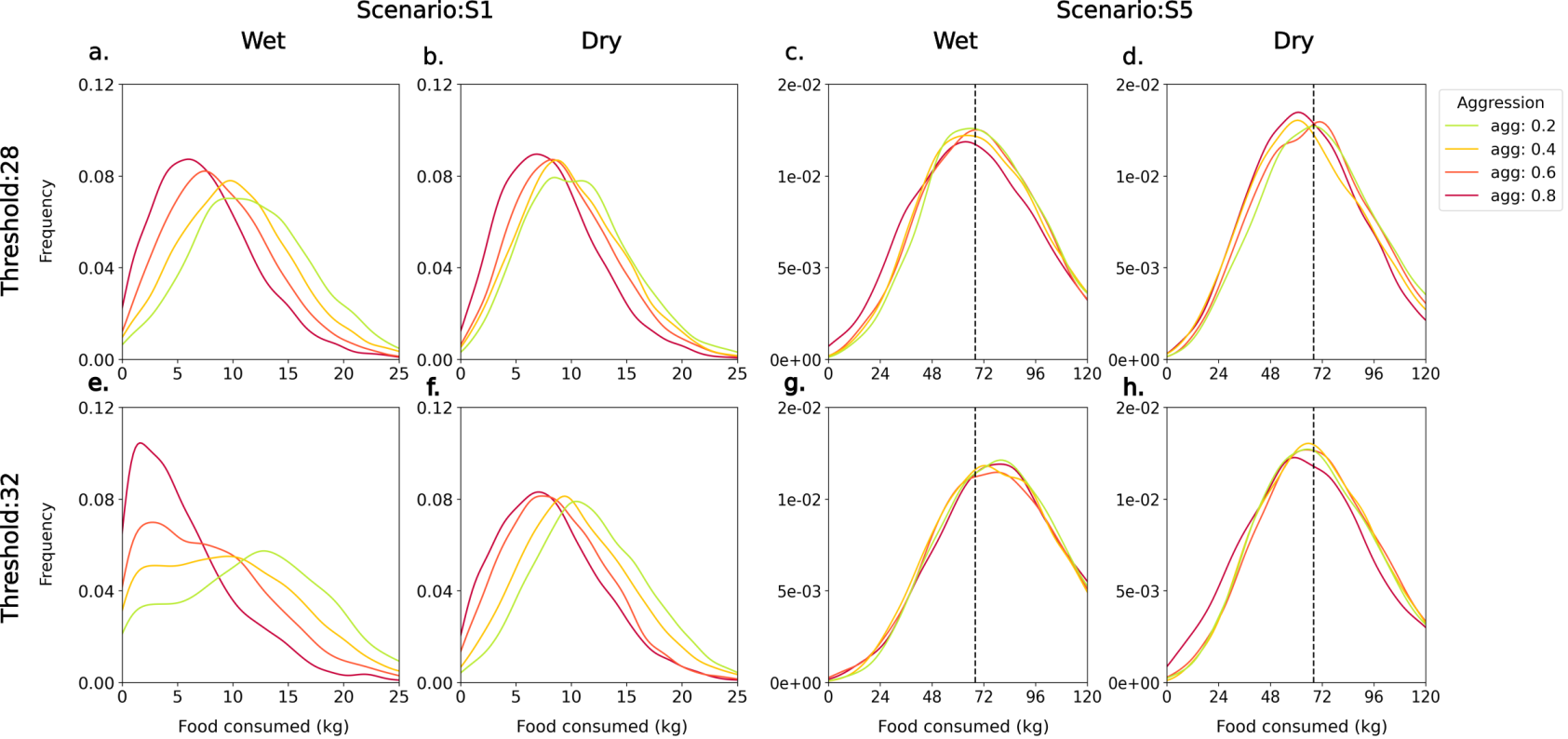}
    \caption{The probability distribution of food consumption from the forest with varying aggression factors.}
    \label{fig:forest_food_consumed}
\end{figure*}

\begin{figure}[h!]
    \centering
    \includegraphics[width=0.625\textwidth]{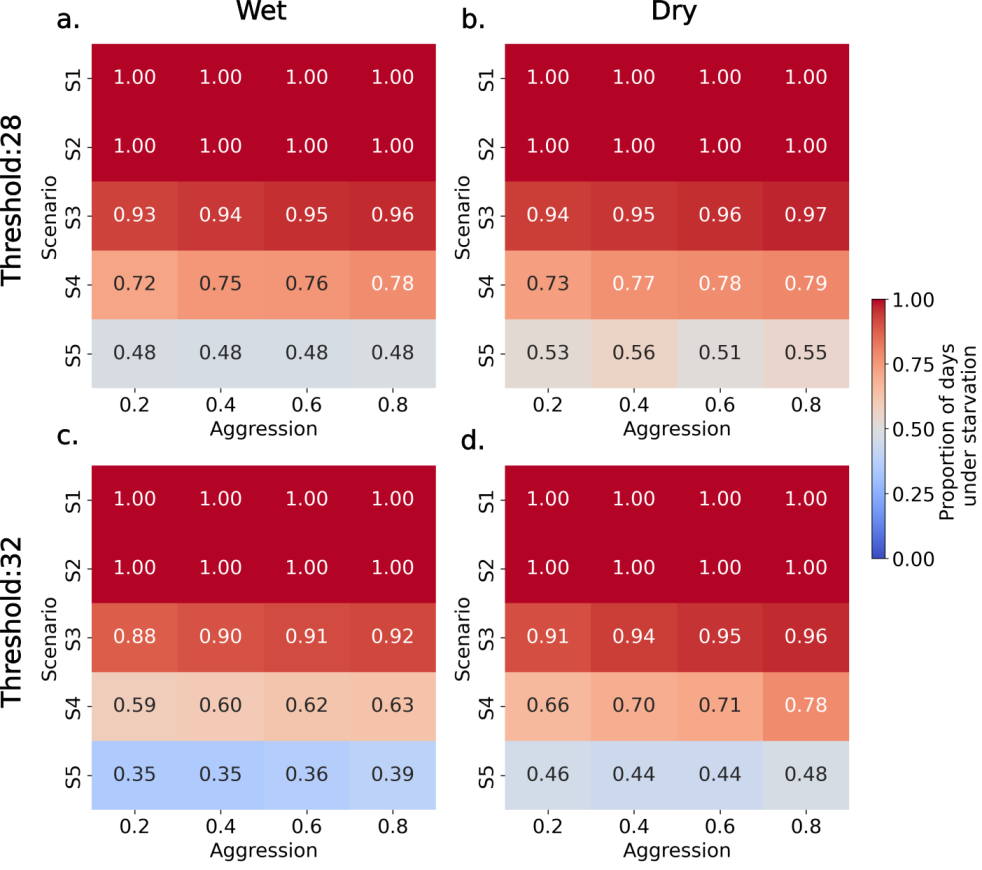}
    \caption{The likelihood of starvation considering all simulated daily trajectories and only food consumed from the forest.}
    \label{fig:prop_days_starved_forest}
\end{figure}

As the elephant agent's initial position in each simulation is located within the forest region, the evolution of the trajectories is based on diverse combinations of agent states and environmental factors. In particular, not all simulated trajectories lead to crop-raiding incidents; rather, only a subset of trajectories traverse mountainous valleys to reach croplands for raiding activities. The significance of these simulations lies in identifying scenarios where crop raiding is prevalent, thus indicating increased human-elephant conflicts within the study region. The proportion of trajectories that reach croplands could be used to calculate the probability of crop raids, with higher proportions suggesting greater risk (Fig.~\ref{fig:prop_traj_raiding}). Specifically, we define scenarios where more than 80\% of the simulated trajectories engage in crop raids as the most critical for human-elephant conflicts. Fig.~\ref{fig:prop_traj_raiding} (c) reveals the situations that are the most critical. 

A combination of three factors leads to at least 80\% of trajectories with crop raiding incidents: (i) agents are not in \textit{thermoregulation} mode, (ii) food availability is limited in the forest, and (iii) aggression is high. The first factor is observed during the wet months and higher thermoregulation thresholds ($T_{threshold}$=32$^\circ$C), during which the agents allocate their daily activity budget solely to \textit{random-walk}, and \textit{foraging} modes, thereby allowing them to venture into the croplands. The second and third factors are simulated by changing food availability (S1-S3 scenarios) and aggression levels. When food is very scarce (Scenario S1), all aggression levels make the situation more critical for human-elephant conflicts. When more food was available in the forest (scenario S2), only higher aggression levels (0.6 and 0.8) resulted in critical situations. With even more food availability (scenario S3), only the highest aggression level (0.8) resulted in a critical situation. 



One of the most challenging concerns is the recurrence of raids exhibited as multiple visits by the same crop raiding elephant agent. Fig. \ref{fig:prop_reentry} shows the probability of making multiple returns to plantations by crop raiding elephant agents. As expected, the likelihood of individuals returning to plantations increases with the level of aggression and under conditions of food scarcity. However, the agent's recurrent visit to the plantations is also influenced by the need for thermoregulation. Figs.~\ref{fig:prop_traj_raiding} and \ref{fig:prop_reentry} illustrate the trade-off between \textit{thermoregulation} and \textit{foraging} in order to optimize overall fitness under varying circumstances.



The proportion of crop raid episodes in which elephants are in a food-deprived state is given in Fig. \ref{fig:prop_incidents_deprecated}. The remaining episodes are crop raid incidents where elephants have become habituated to crops.

To investigate how agents meet their dietary needs within the forest, we examined food consumption in various scenarios of food availability within the forest. Fig.~\ref{fig:food_distributions} illustrates the daily total food consumption (from forests and croplands) of the elephant agents during the dry and wet months in various scenarios for the lowest aggression level (0.2). With 192 trajectories per month, there are 1536 simulated trajectories for wet months (8 months total) and 768 for dry months (4 months total). For comparative analysis, we sampled an equal number of days (5000 days) from the simulated trajectories for the wet and dry months to generate Fig.~\ref{fig:food_distributions}. The horizontal dotted red line in Fig.~\ref{fig:food_distributions} represents the threshold of 68 $kg$, which corresponds to the daily dietary requirement of the elephant agents within the simulations. The agents met their daily nutritional needs when food consumption exceeded this threshold. However, if consumption falls below this threshold, the agent has not consumed enough food and is considered to be in a state of starvation.

We also investigated daily food consumption by elephant agents only from the forest at various levels of aggression and in different scenarios of forest food availability, for dry and wet months. Fig.~\ref{fig:forest_food_consumed} shows the probability density of food consumption only from the forest. Furthermore, we computed the probability of starvation, representing the probability of consuming less than the agent's dietary requirement of 68 $kg$. This probability is calculated as the area under the probability density curve to the left of the 68 $kg$ threshold (Fig.~\ref{fig:forest_food_consumed}). The probability of starvation when eating food only from the forest is presented in Fig.~\ref{fig:prop_days_starved_forest}. As aggression levels increase, foraging efficiency in forests decreases, as agents are forced to prioritize movement toward croplands for raiding. This compromise results in suboptimal foraging within forests, especially in scenarios of food scarcity (Scenario: S1-S4), as seen in Fig. \ref{fig:forest_food_consumed} (a), (b), (e), and (f). In scenarios of high food availability (Scenario: S5), the influence of aggression is less significant and there is no discernible variance in food consumption with different levels of aggression (Fig. \ref{fig:forest_food_consumed} (c), (d), (e), and (h)).

\section{Discussion}\label{Discussion}

Numerous conflicts occur between humans and elephants in the peripheries of forests in the Western Ghats. For example, the Chinnakanal and Santhanpara regions in the Idukki district of Kerala, India, often experience problems with wild elephants \citep{keralakaumudi_chakkakomban,manorama_news_chakkakomban,arikkomban_news_Shaji}. Two of the most well-known tuskers from this area, infamous for raiding and encroaching on residential areas, are named Arikomban and Chakkakomban. Arikomban, known for his frequent raids of rice stores, was sedated, captured, and moved to Periyar Wildlife Reserve in Kerala and then to Kalakkad Mundanthurai Tiger Reserve in Tamil Nadu after multiple efforts by forest authorities to prevent his intrusions \citep{mathrubhumi_news_arikkomban}. Chakkakomban, known for his love of jackfruit, persistently causes problems for farmers, causing many to transition to different crops or to harvest jackfruits prematurely to minimize losses \citep{arikkomban_news_Shaji}. Although considerable debate continues about the underlying causes of conflict, some attribute it to food scarcity within forests \citep{arikkomban_news_Shaji}, forest fragmentation \citep{arikkomban_news_Perinchery}, and food habituation \citep{keralakaumudi_chakkakomban}. Previous studies using ABMs have explored food or resource scarcity as an important driver of conflict \citep{Mamboleo2021}. It is common to believe that crop raids occur during times of food scarcity, such as in the dry months. However, recent studies in South India have revealed a surprising pattern in which vulnerability to crop raids increases during the wet months \citep{bal_2011,gubbi_2012,Anoop_2023}. In addition, recent studies have hypothesized that the high nutritional quality of crops attracts elephants outweighing the human-induced stress response associated with crop raid events \cite{Pokharel2019}. The key takeaway from these studies is that the conditioning of elephants toward crops is an important driver of conflict. Furthermore, a recent survey on human-elephant conflict in Wayanad district in Kerala, India revealed the changing aggression behavior in elephants, with 97\% of the respondents reporting increased elephant aggression \citep{Anoop_2023}. In particular, all participants agreed that elephants previously exhibited avoidance behaviors in response to human deterrents, such as torches, firecrackers, and drums, which are no longer effective. Several social, economic, and cultural variables are critically important determinants of elephant-human relations. For example, for the Gudalur landscape in southern India, ethnicity and shared histories of living are important to explain aspects such as tolerance and coexistence in a crop-raiding scenario \citep{Thekaekara_2021}. These findings suggested that further research is needed to examine and create targeted approaches to address the behavioral and cognitive aspects of elephants. Our prototype ABM serves as a starting point for exploring these complex interactions with appropriate sub-models and simulation experiments.




Previous research has investigated how climatic variables, such as precipitation and water availability, affect the spatial behavior of elephant populations \citep{Sukumar1989}. It was identified that among all climatic factors, the availability of water, including its direct access to consumption and its influence on vegetation, plays the most crucial role in shaping the seasonal cycles of elephants. Elephants were seen to gather in high density near river valleys during the dry months and then disperse across their habitats during the wet months. Our model reproduced similar trends in both dry and wet seasons and examined how these, along with other variables, could influence conflict patterns. In our investigation, Conflict Cluster II emerged solely under one of the experimental conditions, and the availability of water was identified as a key factor in its emergence. However, the impact of thermoregulation on elephant behavior patterns and its consequent effect on crop raiding incidents has not been extensively investigated in prior studies, along with variables such as crop habituation and aggression. 

The weather patterns and topography of the Wayanad forest range are similar to those of Seethathode, as both are located in the Western Ghats. Both areas experience a southwest monsoon from June to September and a northeast monsoon from November to December. During the southwest monsoon, there is heavy rainfall, while the period from February to May is characterized by dry conditions. However, the main difference between the two areas is the dominant crop species. In Wayanad, coffee and paddy are the main crops, while Seethathode is currently dominated by rubber cultivation. The Wayanad region exhibits two distinct seasonal peaks in human-elephant conflict: the monsoon season (June-September) coinciding with jackfruit ripening and the paddy harvest period (October-December). The reported conflict arises when there is an abundance of high-quality forest forage during the monsoon season. This suggests that food scarcity is not the primary reason why elephants are attracted to croplands. Instead, they are drawn mainly to croplands due to the presence of mature jackfruit trees and paddy. In addition, heavy monsoon rains pose challenges to effective crop guarding, potentially increasing the risk of crop raiding during this time. These observations suggest that seasonal resource availability and human activity patterns play a crucial role in shaping human-elephant conflict dynamics within the region. 

Another separate study conducted in Karnataka's Kodagu district, which shares similar climatic conditions with the Wayanad and Seethathode regions, revealed similar conflict patterns with increased activity during the monsoon and postmonsoon seasons \citep{bal_2011}. Perennial crops such as bananas, coconuts, and arecanuts, which carry fruits throughout the year, have been reported to be vulnerable to raids during the monsoon and post-monsoon seasons, even when there is abundant forage available within the forests. This study highlights the need for more studies on finer spatial scales to understand the influence of rainfall on the patterns of crop raiding exhibited by elephants. 

To effectively address human-wildlife conflict, it is crucial to have a comprehensive understanding of the root causes and to go beyond generic solutions that may not be suitable for every situation. The problem of moving beyond one-size-fits-all solutions is addressed in a study conducted in Nagarahole National Park \citep{gubbi_2012}. Although generic solutions such as the creation of water holes or the cultivation of feed can be suggested, the effectiveness of these solutions is questioned in the study mentioned above, as the habituation of the crop by elephants emerges as the main driver. In addition, the efficiency of physical barriers, such as electric fences, depends on careful maintenance and surveillance, emphasizing the importance of evaluating each situation individually rather than making general recommendations.

\subsection{Limitation} \label{sec:limitation}
In conflict scenarios, the behavior of humans and elephants is characterized by complexity and uncertainty. The drivers of crop raid are also complex and multifaceted. Our modeling exercise is only a first step in working within such constraints. Thus, there are several limitations of the prototype ABM. We list a few major limitations below.

The elephant movement data used for movement model calibration comes from a different region due to the lack of specific data in the Periyar-Agastyamalai region. In addition, in cases where specific information on Asian elephants was lacking, we used insights from African elephants as a substitute. We employ probabilistic models to account for uncertainty to enhance the robustness of our simulations. Our current implementation utilizes a single agent, which may not fully represent the complexities of elephant social structures and interactions. However, the ABM framework is flexible and capable of accommodating matriarchal herds in the next version with new agents with associated behavioral submodels. Furthermore, with the appropriate data and parameterization, the present prototype ABM can be adapted for more nuanced simulations that account for group dynamics. Another limitation is the simplification of representing food resources. We assume a simple Bernoulli distribution of forest and plantation cells within the land use classes identified according to the LULC map and questionnaire data; however, in reality, factors such as vegetation density, water sources, and soil quality contribute to a heterogeneous distribution of resources. Another limitation of the current prototype ABM model is that migratory and long-term mobility patterns of elephants are not included. Each simulation captures only short-term behavior, focusing on the movement dynamics of elephants when they are in close proximity to plantations. Even with these limitations, the insights obtained are a valuable first step towards building a full fledged model in the Periyar-Agasthyamali complex.


\section{Conclusions}\label{sec:Conclusions}

Understanding the dynamics driving elephant space use is crucial for developing effective strategies to mitigate human-elephant conflicts and ensure co-existence. The agent-based modeling framework allows the incorporation of observational datasets and ecological understanding with a natural uncertainty quantification process (due to the probabilistic nature of the simulation) to study emergent patterns. Our prototype ABM is the first to incorporate the interaction between food scarcity, habituation, thermoregulation, and aggression levels to study the emerging movement patterns of Asian elephants and the corresponding crop-raiding behaviors. The main challenges of lack of specific data to develop spatially explicit agent-based models were addressed through a careful mathematical modeling approach to arrive at movement and cognition models for solitary bull elephants in the Periyar-Agasthyamali complex. A questionnaire-based survey dataset collected in the region by one of the coauthors was used to inform the modeling choices and as a validation source. Relocation data from radio-tagged Asian elephants collected by one of the coauthors in Indonesia was used as a proxy together with a calibration process to adapt it to Asian elephants in the Periyar-Agastyamali complex. 

This prototype ABM can be used to provide qualitative and quantitative data to analyze "what if" scenarios for mitigation, as suggested in the recent survey \citep{Anoop_2023}. The current model serves as a foundational step, paving the way for incorporating more complexities to create a data-informed, ecology-based decision-support system for wildlife management.

\subsection{Future Directions}
The current prototype ABM can be used to further explore the intricacies of human-elephant conflict. A key study would be to investigate the impact of water availability on space use and emerging conflict scenarios. Furthermore, the model can be used to explore the potential to serve as an effective tool for forecasting the possible evolution of human-elephant conflict in a changing climate. The present model uses rule-based cognition submodels, but more sophisticated machine learning techniques could be used to develop submodels with more capabilities. For example, sequential learning models could allow agents to learn from past experiences and adapt their strategies, while reinforcement learning could help uncover the best foraging and raiding strategies in simulated settings. Furthermore, human-elephant conflict represents a multifaceted competition for resources, and employing a game-theoretic framework provides a promising approach to exploring solutions. These future research directions ultimately aim to develop intelligent decision support tools that identify practical and satisficing solutions to ensure the well-being of humans and elephants.


\appendix
\section*{Appendix}
\section{Data availability} \label{sec:dataAvailability}

\begin{enumerate}
\item We utilise the data from a previous study (Sect.~\ref{sec:field_survey_and_data_collection}) \cite{Oommen_2017_a}. Field data collection on human subjects included detailed questionnaire surveys involving 428 households as well as oral histories and key informant interviews related to the dynamics of conflict along the forest fringe.  All interviews with human subjects in the study area are covered under an ethics approval to the co-author (Oommen) from the University of Technology Sydney (UTS HREC 2012 183A). We have provided a spreadsheet with the relevant field data based on the questionnaire survey as well as the GPS locations (acquired using handheld GPS units) related to land use in landholdings (\url{https://github.com/quest-lab-iisc/abm-elephant-project/blob/master/data/seethathodu_data.xlsx}). The names and addresses of individual respondents have been anonymized following guidelines as per the ethics approval.

\item The elephant movement data (Sect. \ref{sec:calibrating_HMM_parameters}, movebank study name: Elephants Java FZG MPIAB DAMN) can be found here: \url{https://www.movebank.org/cms/webapp?gwt_fragment=page=studies,path=study56232621}. 

\item Our ABM can be accessed on Github at \url{https://github.com/quest-lab-iisc/abm-elephant-project}. 

\end{enumerate}

\section{Classification of agricultural plots within the study area}\label{sec:classification_agricultural_plots}

Classification of agricultural plots within the study area based on the proportion of the rubber canopy and home gardens:  \\
1. None: No rubber canopy or HG present \\
2. HG: No rubber canopy and 100\% of land used as home garden \\
3. 0-25rubberXHG: $0\% < rubber-canopy <= 25\%$ and remaining area is used as home garden \\
4. 25-50rubberXHG: $25\% < rubber-canopy <= 50\%$ and remaining area is used as home garden \\
5. 50-75rubberXHG: $50\% < rubber-canopy <= 75\%$ and remaining area is used as home garden \\
6. 75-100rubberXHG: $75\% < rubber-canopy <= 100\%$ and remaining area is used as home garden \\
7. 100rubber: $100\%$ area is occupied by rubber canopy and no home garden. \\

\begin{figure}[h!]
    \centering
    \includegraphics[width=0.45\textwidth]{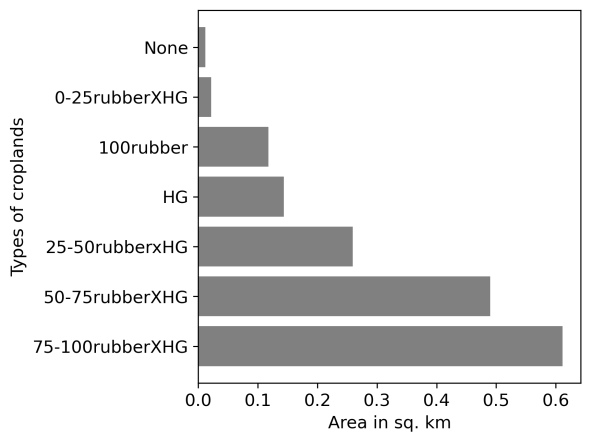}
    \caption{The types of agricultural plots within the study area as informed from the field survey}
    \label{fig:agricultural-plot-distribution}
\end{figure}

\section{Convergence analysis}\label{Convergence_analysis}

To ensure convergence in simulations, it is necessary to assess the variance in the model's output \citep{Lee2015}. This is crucial in simulation-based experiments like agent-based models, where statistical averages from limited repetitions are considered. Computational cost limits the number of repetitions without compromising model accuracy. The Lorscheid method, which compares the convergence in mean and variance of the model's output, was used to determine the appropriate number of repetitions. This method uses the coefficient of variation, $CV = \frac{\sigma}{\mu}$, where $\mu$ and $\sigma$ are the sample mean and standard deviation. This metric was selected because it does not assume a normal distribution. The minimum number of repetitions is determined using a threshold $\epsilon$, representing the point where the difference in the coefficient of variation stays within $\epsilon$ \citep{Venugopalan2023}. As repetitions increase, the difference between sample and population metrics decreases, making them more similar. Experiments on multiple outputs and sample sizes identified the smallest sample size ($n_{min}$) as the point where the estimated CV difference does not deviate significantly.

\begin{figure}[h!]
    \centering
    \includegraphics[width=0.65\textwidth]{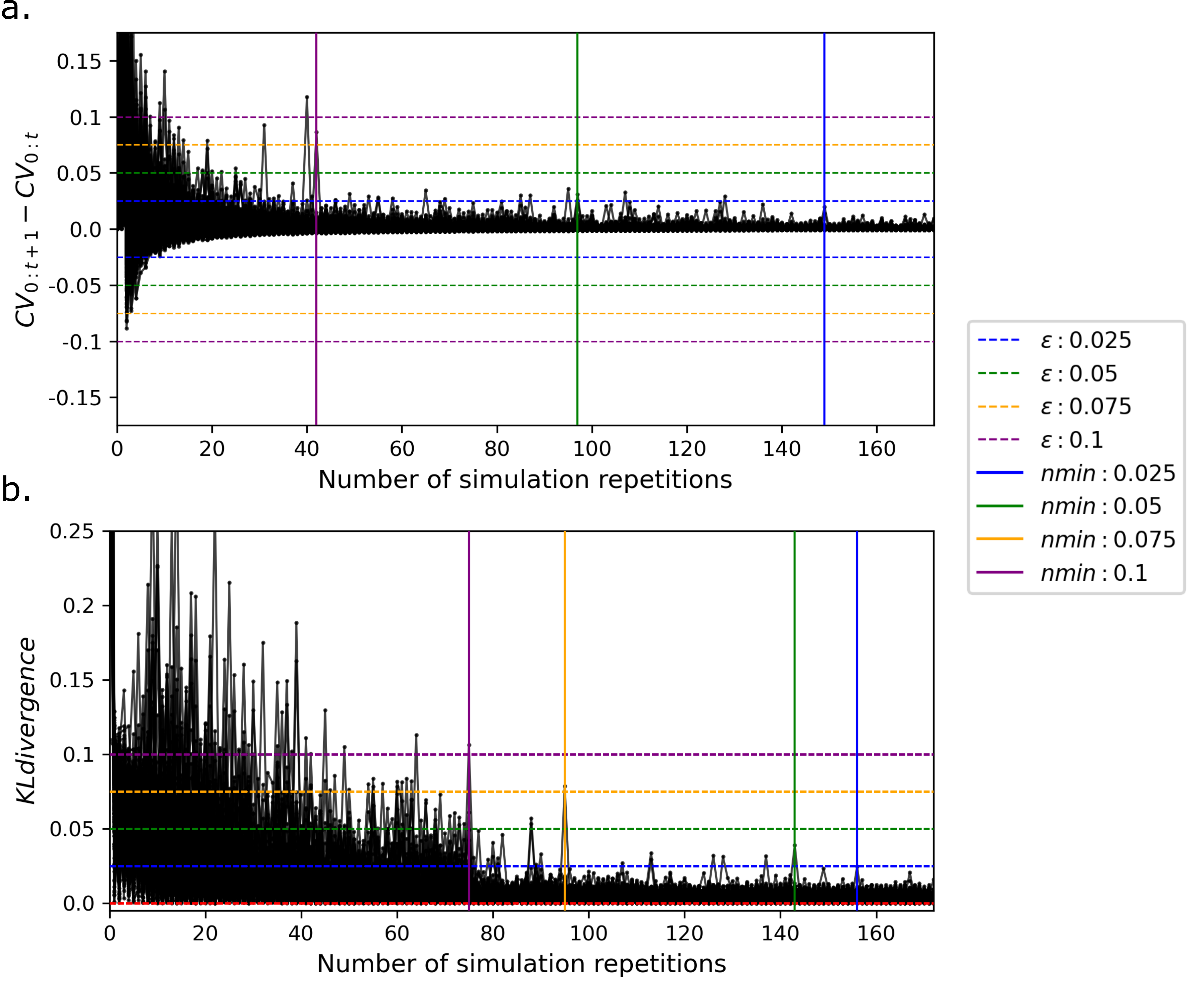}
    \caption{(a) The variation in the coefficient of variation as the number of simulation repeats increases. The model outputs being analyzed are the estimated space usage using the MCP and the 95\% KDE. (b) The KL divergence of the 2D agent distribution as the number of simulation repeats increases. Both plots represent 500 different random model parameterizations. The horizontal lines represent the threshold levels, while the vertical lines indicate the simulation repeats at which the change remains within the threshold.}
    \label{fig:coefficient-of-variation}
\end{figure}

The elephant agent's trajectories displayed a significant amount of variability. The model outputs used for the convergence analysis were the MCP and 95\% KDE measurements of the elephant agents' space utilization. The 95\% KDE provided an estimate of the core area usage, while the MCP indicated the dispersion of the trajectories from the initial location. Random initialization was chosen to configure the model. The outputs were compared using 500 different model parameterizations and four values of $\epsilon$ (0.025, 0.05, 0.075, and 0.1) (see Fig. \ref{fig:coefficient-of-variation} (a)). Some model parameterizations required more iterations to achieve convergence due to the uncertainty in the trajectory evolution. The values of $nmin$ for each $\epsilon$ value are summarized in Table \ref{tab:number-of-simulations}.

Additionally, the convergence in the spatial arrangement of agents' paths was also assessed by calculating the Kullback-Leibler (KL) divergence in the 2D probability distribution of the agents. Similar to the previous configuration, 500 distinct random parameterizations were taken into account, and various thresholds were employed to determine the convergence (refer to Fig. \ref{fig:coefficient-of-variation} (b)).

Based on the findings of the two convergence studies, a value of 192 was selected as the minimum number of iterations ($nmin$) for subsequent experiments. The compute nodes in the \textit{parampravega} system which was used to run the experiments have 48 cores, and a total of 4 nodes were utilized in each experimental run. It was determined that the chosen $nmin$ value of 192 successfully achieved the required convergence in all subsequent experiments and analysis.

\begin{table*}[h!]
 \caption{The minimum number of repeats for convergence}\label{tab:number-of-simulations}
 \begin{tabular}{p{1cm}p{3cm}p{3cm}}
 \toprule
 \textbf{$\epsilon$} & \makecell{\textbf{$nmin$} \\ (estimated using CV)}  & \makecell{\textbf{$nmin$} \\ (estimated using KL\\ divergence)}\\
 \midrule
   0.1 & 42 & 75\\
   0.075 & 42 & 95\\
   0.05 & 94 & 143\\
   0.025 & 149 & 156\\
 \bottomrule
 \end{tabular}
\end{table*}

\section{Model verification: movement of elephants taking into account the topography and slope}\label{appendix:day_night_trajectory}

\begin{figure*}[h!]
    \centering
    \includegraphics[width=0.725\textwidth]{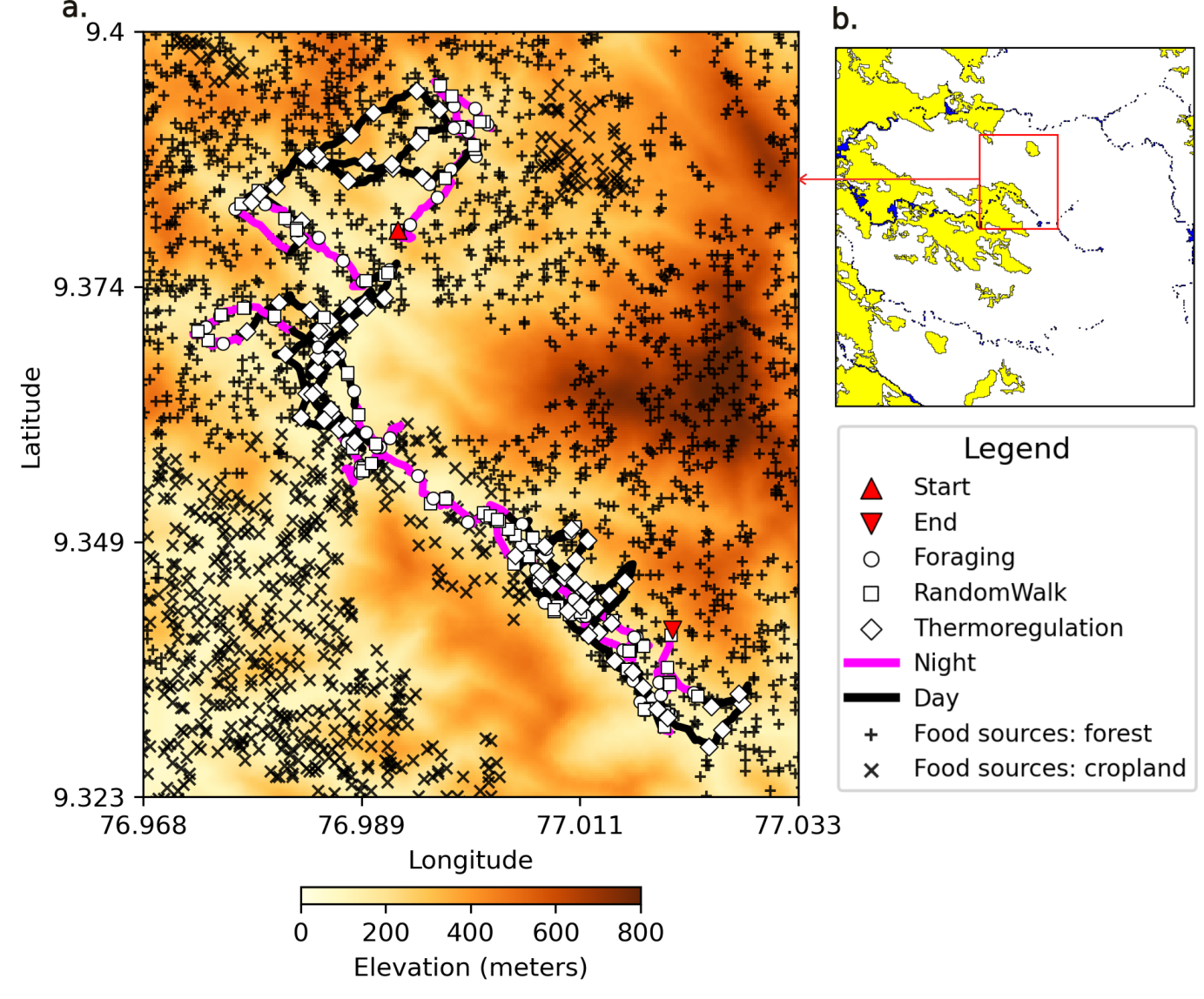}
    \caption{(a) A simulated elephant trajectory sequence that highlights the activity during the day and night shown against the DEM. The figure illustrates how elephant agents navigate mountainous terrain using valleys and flatter regions. (b) The simulation extent is displayed, with the zoomed in region in part (a) highlighted by the red box. Note that the trajectory sequence shown is the same as in Fig.~\ref{fig:day_night_trajectories}.}
    \label{fig:day_night_trajectories_DEM}
\end{figure*}

Fig.~\ref{fig:day_night_trajectories_DEM} shows a sample trajectory sequence, highlighting the movement patterns of the elephant agents during the day and night against the DEM of the study area. The figure shows how agents navigate through valleys and flatter regions to traverse the mountainous landscape. 

\paragraph{Declaration of Competing Interest} 

The authors state that they do not have any known financial conflicts or personal relationships that might seem to affect the findings presented in this document.

\paragraph{Acknowledgments}
The authors extend heartfelt gratitude to Sanjeeta Sharma Pokharel and Nachiketha Sharma for their insightful discussions on Asian elephant ecology, which have immensely enriched this work. The authors thank Prof. Kartik Shanker (Center of Ecological Sciences, Indian Institute of Science) for discussion and research direction. This work was partially supported by the INSPIRE Faculty Research Grant to DNS from the Department of Science and Technology, Government of India. We are grateful for the fellowship support to AP from the Ministry of Education, Government of India.

\printcredits

\bibliographystyle{elsarticle-num}

\bibliography{refs}

\bio{}
\endbio

\bio{}
\endbio

\end{document}